%% file: main.tex
\documentclass[conference,compsoc]{IEEEtran}
\input{headers/math_commands.tex}

\input{headers/header-lqa}

\input{headers/header}
\input{headers/header-lqa-tail}

\usepackage{hyperref}
\usepackage{url}

\ifCLASSOPTIONcompsoc
  \usepackage[nocompress]{cite}
\else
  \usepackage{cite}
\fi
\usepackage{stfloats}
\usepackage{caption}

\begin{document}
\title{TIGER: Inverting Transformer Gradients via Embedding-Subspace Distance Optimization}




\author{\IEEEauthorblockN{William Kalikman\IEEEauthorrefmark{1},
Ivo Petrov\IEEEauthorrefmark{2},
Dimitar I. Dimitrov\IEEEauthorrefmark{2} and
Martin Vechev\IEEEauthorrefmark{1}\IEEEauthorrefmark{2}}
\IEEEauthorblockA{\IEEEauthorrefmark{1}ETH Zürich}
\IEEEauthorblockA{\IEEEauthorrefmark{2}INSAIT, Sofia University "St. Kliment Ohridski"}
}


\maketitle

\input{paper_files/abstract.tex}

%
\IEEEpeerreviewmaketitle

\input{paper_files/introduction.tex}

\input{paper_files/related.tex}
\input{paper_files/background.tex}
\input{paper_files/methodology.tex}
\input{paper_files/results.tex}
\input{paper_files/limitations.tex}
\input{paper_files/conclusion.tex}
\input{paper_files/acknowledgements.tex}





\bibliography{references}
\bibliographystyle{IEEEtran}

\input{paper_files/appendix.tex}

\end{document}

%% file: headers/math_commands.tex
\usepackage{amsmath,amsfonts,bm}

\def\eqref#1{equation~\ref{#1}}

\def\1{\bm{1}}

\DeclareMathAlphabet{\mathsfit}{\encodingdefault}{\sfdefault}{m}{sl}
\SetMathAlphabet{\mathsfit}{bold}{\encodingdefault}{\sfdefault}{bx}{n}



%% file: headers/header-lqa.tex
\usepackage[utf8]{inputenc} 
\usepackage{lipsum} 

\usepackage[T1]{fontenc}

\usepackage{etoolbox}
\newbool{includeappendix}
\setbool{includeappendix}{true}

\input{headers/lqa/overfull}


\input{headers/lqa/colors}

\input{headers/lqa/listing}

%% file: headers/lqa/overfull.tex
\ifdefined\isoverfull
\else
\fi

%% file: headers/lqa/colors.tex
\usepackage{xcolor} 

\definecolor{my-full-blue}{HTML}{1F77B4}

\definecolor{my-full-orange}{HTML}{FF7F0E}

\definecolor{my-full-green}{HTML}{2CA02C}

\definecolor{my-full-red}{HTML}{d62728}

\definecolor{my-full-purple}{HTML}{9467bd}

\colorlet{my-blue}{my-full-blue!30}
\colorlet{my-orange}{my-full-orange!30}
\colorlet{my-green}{my-full-green!30}
\colorlet{my-red}{my-full-red!30}
\colorlet{my-purple}{my-full-purple!30}


%% file: headers/lqa/listing.tex
\usepackage{listings}

\usepackage{textcomp}

\usepackage{xcolor}

\usepackage[scaled=0.8]{beramono}

\definecolor{ckeyword}{HTML}{7F0055}
\definecolor{ccomment}{HTML}{3F7F5F}
\definecolor{cstring}{HTML}{2A0099}

\lstdefinestyle{numbers}{
	numbers=left,
	framexleftmargin=20pt,
	numberstyle=\tiny,
	firstnumber=auto,
	numbersep=1em,
	xleftmargin=2em
}

\lstdefinestyle{layout}{
	frame=none,
	captionpos=b,
}

\lstdefinestyle{comment-style}{
	morecomment=[l]//,
	morecomment=[s]{/*}{*/},
	commentstyle={\color{ccomment}\itshape},
}

\lstdefinestyle{string-style}{
	morestring=[b]",%
	morestring=[b]',%
	stringstyle={\color{cstring}},
	showstringspaces=false,%
}

\lstdefinestyle{keyword-style}{
	keywordstyle={\ttfamily\bfseries},
	morekeywords={
		function,
		constructor,
		int,
		bool,
		return,
		returns,
		uint
	},
	morekeywords = [2]{},
	keywordstyle = [2]{\text},
	sensitive=true,
}

\lstdefinestyle{input-encoding}{
	inputencoding=utf8,
	extendedchars=true,
	literate=
	{ℝ}{$\reals$}1%
	{→}{$\rightarrow$}1%
	{α}{$\alpha$}1%
	{β}{$\beta$}1%
	{λ}{$\lambda$}1%
	{θ}{$\theta$}1%
	{ϕ}{$\phi$}1%
}

\lstdefinestyle{escaping}{
	moredelim={**[is][\color{blue}]{\%}{\%}},
	escapechar=|,
	mathescape=true
}

\lstdefinestyle{default-style}{
	basicstyle=\fontencoding{T1}\ttfamily\footnotesize,
	style=numbers,
	style=layout,
	style=comment-style,
	style=string-style,
	style=keyword-style,
	style=input-encoding,
	style=escaping,
	tabsize=2,
	upquote=true
}

\lstdefinelanguage{BASIC}{
	language=C++,
	style=default-style
}[keywords,comments,strings]%

%% file: headers/header.tex
\usepackage{booktabs}       
\usepackage{nicefrac}       
\usepackage{microtype}      
\usepackage[flushleft]{threeparttable}
\usepackage{algorithm}
\usepackage[noend]{algorithmic}

\usepackage{url}
\usepackage{xcolor}
\usepackage{xspace}
\usepackage{etoolbox}
\usepackage{fontawesome}
\usepackage{enumitem}
\usepackage{amsmath,amssymb}
\usepackage{array}
\usepackage{multirow}
\usepackage{wrapfig}
\usepackage{adjustbox,booktabs}
\usepackage{siunitx}
\usepackage{caption}
\usepackage{fontawesome}
\usepackage{stackengine}
\usepackage{svg}
\usepackage{mathtools}
\usepackage{xspace}
\usepackage{wrapfig}
\usepackage{makecell}
\usepackage{graphicx}
\usepackage{booktabs}
\usepackage{longtable}
\usepackage{hyperref}
\usepackage[labelformat=simple]{subcaption}
\usepackage[most]{tcolorbox}
\usepackage{diagbox}
\usepackage{ulem}
\usepackage[capitalize,noabbrev]{cleveref}

\usepackage{colortbl}

\input{headers/lqa/tikz}

\usepackage{amsthm}
\usepackage{stmaryrd}

\theoremstyle{plain}
\newtheorem{theorem}{Theorem}[section]

\theoremstyle{definition}

\theoremstyle{remark}

\theoremstyle{equation_th}

\newcolumntype{x}[2]{S[table-format=#1.#2,table-auto-round]}
\newcolumntype{M}{>{$}l<{$}}

\renewcommand{\emph}[1]{\textit{#1}}

\newcommand{\method}{\textsc{TIGER}\xspace}
\newcommand{\dager}{\textsc{DAGER}\xspace}
\newcommand{\lamp}{\textsc{LAMP}\xspace}

\newcommand{\gemma}{\textsc{Gemma-3-4B-IT}\xspace}
\newcommand{\embgemma}{\textsc{EmbeddingGemma-300M}\xspace}

\newcommand{\gpttwo}{\textsc{GPT-2}\xspace}

\newcommand{\wikitext}{WikiText\xspace}

\newcommand{\RETURN}{\STATE \textbf{return} }
\renewcommand{\algorithmicrequire}{\textbf{Input:}}
\renewcommand{\algorithmicensure}{\textbf{Output:}}

\newcommand{\BatchSize}{B}
\newcommand{\NumBlocks}{L}
\newcommand{\HiddenDim}{d}

\newcommand{\Vocab}{\mathcal{V}}
\newcommand{\VocabSize}{V}

\newcommand{\Embeddings}{\bm{E}}
\newcommand{\TokenEmbedding}[1]{e_{#1}}

\newcommand{\TrainingLoss}{\mathcal{L}}
\newcommand{\SeqLength}[1]{T_{#1}}

\newcommand{\MaxSeqLength}{T_{\text{max}}}
\newcommand{\NumTokens}{T}

\newcommand{\HiddenState}[1]{\bm{Z}^{#1}}

\newcommand{\QueryWeight}[1]{\bm{W}^{#1}_{Q}}
\newcommand{\KeyWeight}[1]{\bm{W}^{#1}_{K}}
\newcommand{\ValueWeight}[1]{\bm{W}^{#1}_{V}}

\newcommand{\QueryActivation}[1]{\bm{Q}^{#1}}
\newcommand{\KeyActivation}[1]{\bm{K}^{#1}}
\newcommand{\ValueActivation}[1]{\bm{V}^{#1}}

\newcommand{\GradQueryActivation}[1]{\frac{\partial \TrainingLoss}{\partial \QueryActivation{#1}}}

\newcommand{\GradQueryWeight}[1]{\frac{\partial \TrainingLoss}{\partial \QueryWeight{#1}}}

\newcommand{\AttentionMask}{\bm{M}}
\newcommand{\Preprocess}{f^0}
\newcommand{\TransformerBlock}[1]{f^{#1}}

\newcommand{\RecoveredInput}[1]{\widehat{\bm{Z}}^{#1}}
\newcommand{\OptimizedInput}[1]{\bar{\bm{Z}}^{#1}}
\newcommand{\GradientSubspace}[1]{\mathcal{S}^{#1}}
\newcommand{\RecoveredSubspace}[1]{\bar{\mathcal{S}}^{#1}}

\newcommand{\SubspaceProjection}[1]{\bm{P}^{#1}}
\newcommand{\SubspaceDedupProjection}[1]{\bm{P}^{#1}_\text{prev}}
\newcommand{\SubspaceBasis}[1]{\bm{U}^{#1}}
\newcommand{\RotatedSubspaceBasis}[1]{{\bm{U}}^{#1}_\text{prev}}
\newcommand{\RecoveredSubspaceProjection}[1]{\bm{\bar{P}}^{#1}}

\newcommand{\NumInitializations}{N_\text{init}}
\newcommand{\NumSteps}{N_\text{steps}}
\newcommand{\MaxLayers}{L_\text{max}}

\newcommand{\ForwardLoss}{\ell_\text{forw}}
\newcommand{\BackwardLoss}{\ell_\text{back}}
\newcommand{\DedupLoss}{\ell_\text{dedup}}
\newcommand{\Loss}{\ell}

\newcommand{\DedupWeightCoeff}{\lambda_\text{dedup}}
\newcommand{\InvJitterCoeff}{\lambda_\text{inv}}
\newcommand{\BackwardCoeff}{\lambda_\text{back}}

\newcommand{\SpanDistance}{\mathcal{D}}

\newcommand{\InitMu}[1]{\bm{\mu}_{#1}}
\newcommand{\InitSigma}[1]{\bm{\Sigma}_{#1}}

\newcommand{\LearningRate}{\eta}

\newcommand{\Token}[1]{t_{#1}}
\newcommand{\RecoveredToken}[1]{\hat{t}_{#1}}

\definecolor{acceptblue}{HTML}{6494EA}
\hypersetup{citecolor=acceptblue}

\definecolor{lightred}{HTML}{ffcbc7}
\definecolor{gemini}{HTML}{4285F4}
\definecolor{claude}{HTML}{f3e9d7}
\definecolor{deepseek}{HTML}{FADA4B}
\definecolor{qwen}{HTML}{FA574B}
\definecolor{oai}{HTML}{10a37f}
\definecolor{LightGreen}{HTML}{CCFFCC}

\lstdefinestyle{mystyle}{
    breaklines=true,
    basicstyle=\scriptsize\ttfamily,
    numbers=none,
    language={},
    framextopmargin=0pt,
    framexbottommargin=0pt,
    breakindent=0pt,
    showspaces = false,
    keywordstyle=\bfseries,
    showstringspaces=false,
    columns=fullflexible,
    morekeywords={Answer}
    moredelim=[**][\bfseries]{!!}
}

\newtcblisting{prompt}[2][]{
    arc=3pt, outer arc=3pt,
    width=\linewidth,
    left=1mm,
    top=0mm,
    bottom=0mm,
    title=#2, 
    colback=lightred!5!white,
    colframe=black,
    fonttitle=\bfseries,
    listing only, 
    listing options={style=mystyle},
    breakable,
    #1
}

\newtcblisting{smallprompt}[2][]{
    arc=3pt, outer arc=3pt,
    width=0.9\linewidth,
    left=1mm,
    top=0mm,
    bottom=0mm,
    title=#2, 
    colback=lightred!5!white,
    colframe=black,
    fonttitle=\bfseries,
    listing only, 
    listing options={style=mystyle},
    breakable,
    #1
}

\newtcblisting{response}[2][]{
    arc=3pt, outer arc=3pt,
    width=0.9\linewidth,
    left=1mm,
    top=0mm,
    bottom=0mm,
    title=#2, 
    colback=lightred!5!white,
    colframe=oai,
    fonttitle=\bfseries,
    listing only, 
    listing options={style=mystyle},
    breakable,
    #1
}

%% file: headers/lqa/tikz.tex
\usepackage{tikz}

\usetikzlibrary{calc,decorations,decorations.pathmorphing}
\usetikzlibrary{positioning,fit,arrows}
\usetikzlibrary{decorations.markings}
\usetikzlibrary{shapes,shapes.geometric}
\usetikzlibrary{shadows,patterns,snakes}
\usetikzlibrary{backgrounds,decorations.pathreplacing,calligraphy,automata}
\usetikzlibrary{intersections}
\usetikzlibrary{angles,quotes}
\usetikzlibrary{plotmarks}
\usetikzlibrary{patterns}
\usetikzlibrary{arrows.meta}
\usetikzlibrary{shapes.misc}
\usetikzlibrary{chains}
\usetikzlibrary{shapes.callouts}

%% file: headers/header-lqa-tail.tex
\input{headers/lqa/references}

%% file: headers/lqa/references.tex
\usepackage[capitalize]{cleveref}

\crefformat{section}{\S#2#1#3}

\crefrangeformat{section}{\S#3#1#4\crefrangeconjunction\S#5#2#6}

\crefmultiformat{section}{\S#2#1#3}{\crefpairconjunction\S#2#1#3}{\crefmiddleconjunction\S#2#1#3}{\creflastconjunction\S#2#1#3}

\newcommand{\crefrangeconjunction}{--}

\crefname{listing}{Lst.}{listings}
\crefname{line}{Lin.}{Lin.}
\crefname{appendix}{App.}{App.}

\newcommand{\app}[1]{%
	\ifbool{includeappendix}{\cref{#1}}{the appendix}%
}
\newcommand{\App}[1]{%
	\ifbool{includeappendix}{\cref{#1}}{The appendix}%
}

%% file: paper_files/abstract.tex
\begin{abstract}
  Federated learning allows multiple clients to jointly train a shared model by sending gradient updates to a central server while keeping raw inputs local. However, prior gradient inversion attacks show that these updates can reveal enough information to reconstruct client inputs. Existing attacks on transformers either optimize dummy inputs to match the true client updates, which is costly and unstable for modern models, or exploit the low rank of attention gradients to identify a subspace containing the true layer embeddings, followed by a discrete membership test for candidate tokens. However, this token test is brittle under numerical noise, i.e., from quantization or Differential Privacy (DP), and scales poorly for encoder models with non-causal attention. We introduce \method{}, a continuous gradient inversion attack that turns this subspace signal into a differentiable objective. Instead of searching over tokens or matching full gradients, \method{} directly optimizes token embeddings to minimize their distance to the subspace. Our experiments demonstrate that on encoder-only models, \method{} substantially improves both reconstruction quality and runtime over existing attacks, while on decoder models, \method{} is more robust than prior subspace-based attacks, enabling the first successful reconstructions in DP-defended federated learning settings.
\end{abstract}

%% file: paper_files/introduction.tex
\section{Introduction}

Federated Learning (FL) has emerged as a practical paradigm for collaboratively training and adapting machine-learning models over decentralized data. In FL, clients keep their raw data local and communicate only model updates, such as gradients or parameter deltas, to a central server that aggregates them to improve a shared model~\cite{federated}. This is particularly appealing for large language models (LLMs), whose training requires large amounts of data, and whose adaptation often requires domain-specific information that may be sensitive or proprietary~\cite{fatellm,chenfederated,yaofederatedllm,jiangfederatedllm}. The privacy requirement is particularly strong in privacy-critical domains such as medicine~\cite{manoelmedical,zhanghealthcare,fedmedlora,sadilekhealth}, law~\cite{fedlegal,fedjudge}, and finance~\cite{flowertune,byrddp}, where centralizing training data may be infeasible due to confidentiality requirements, regulatory constraints, or institutional data regulations.

\subsection{Gradient Inversion Attacks} However, keeping raw data with the client does not by itself guarantee privacy. Gradient inversion attacks have shown that the model updates shared in FL can contain enough information to reconstruct the client data used to compute them~\cite{dlg,idlg,geiping}. These attacks have been highly successful on image data~\cite{dlg,idlg,geiping,gradinversion,generativegradinversion,gradvit,cocktailparty}, and subsequent work has extended them to the text domain~\cite{tag,lamp,film,poolerattack,dager}. Textual reconstruction, however, remains challenging on modern LLMs. Existing attacks often rely on expensive gradient-matching optimization~\cite{dlg,tag,lamp}, strong language-model priors~\cite{lamp,film}, or discrete filtering procedures that are sensitive to numerical perturbations~\cite{dager}. These limitations become especially important in realistic FL deployments, where clients may train with nontrivial batch sizes, and where communicated gradients may be quantized or perturbed with noise. As a result, the apparent failure of existing attacks in such settings may underestimate the real privacy risks of gradient inversion attacks.

\begin{figure*}[!t]
    \centering
    \includegraphics[width=\textwidth]{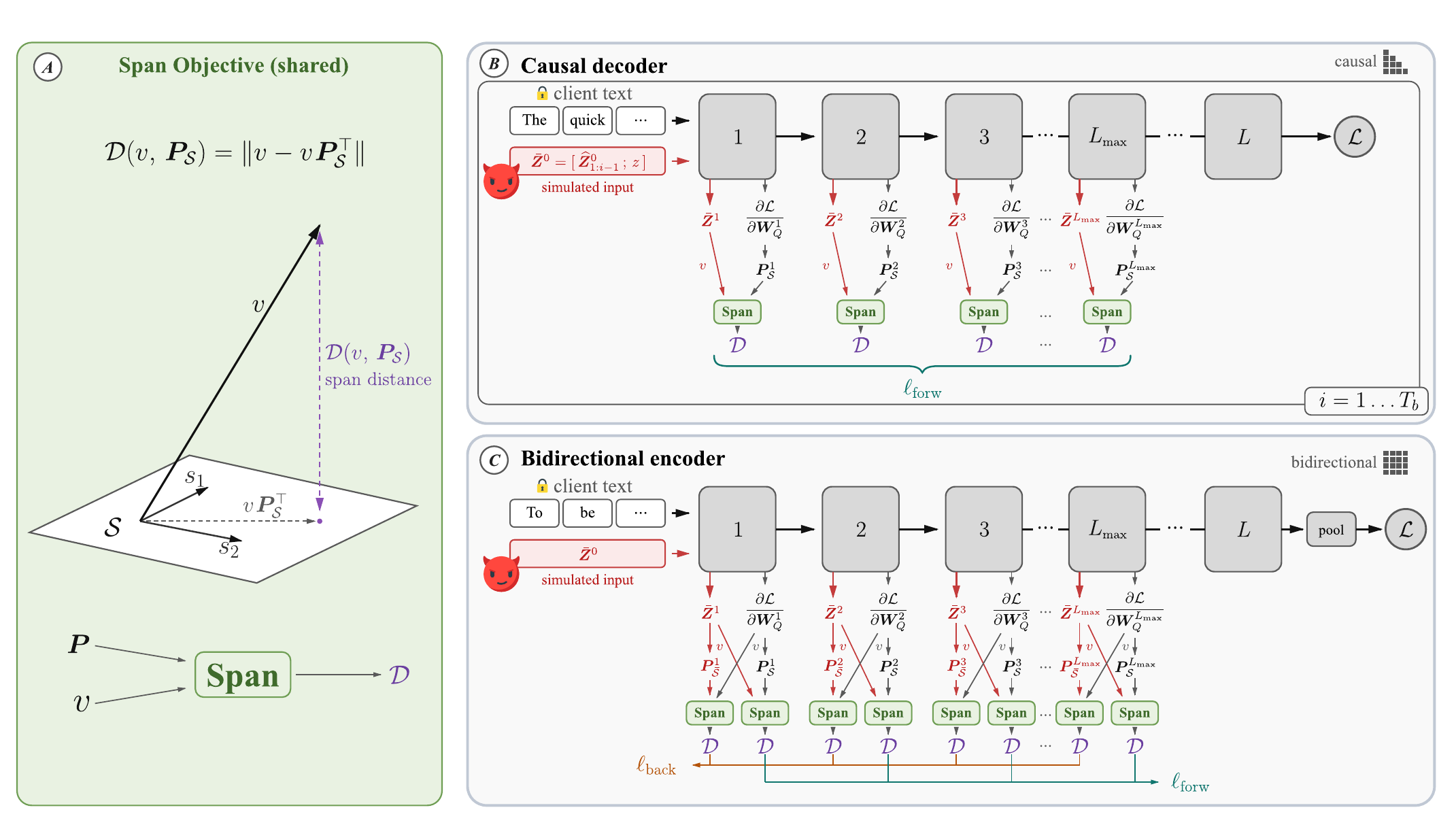}
    \caption{Overview of \method{}. \textbf{(a)}~The shared span objective: the span distance $\SpanDistance(v,\,\SubspaceProjection{}_{\GradientSubspace{}})$ between a vector $v$ and a subspace $\GradientSubspace{}$ with projection matrix $\SubspaceProjection{}_{\GradientSubspace{}}$, computed by each \textsc{Span} block. \textbf{(b)}~Decoder attack: embeddings are recovered position by position ($i = 1 \ldots \SeqLength{b}$) using the forward loss $\ForwardLoss$ only. Red marks attacker-simulated quantities. \textbf{(c)}~Encoder attack: the simulated input $\OptimizedInput{0}$ is optimized so that at every layer $l \le \MaxLayers$ the hidden states $\OptimizedInput{l}$ lie in the gradient-induced subspace ($\ForwardLoss$) and the gradient directions lie in the span of the hidden states ($\BackwardLoss$).}
    \label{fig:overview}
\end{figure*}

\subsection{This work: Continuous Recovery of Text under Large Batches in Defended Settings} To address these limitations, we develop \method{} (\textbf{T}ransformer Data \textbf{I}nversion from \textbf{G}radients via \textbf{E}mbedding-subspace \textbf{R}econstruction), a novel gradient inversion attack that reconstructs private training text from transformer LLM updates. \method{} provides a continuous recovery framework, which focuses on improving robustness in the settings that arise in practical federated learning. Across larger client batches and numerically perturbed gradients from low-precision gradient calculation or DP-style noise, \method{} recovers substantially more text than prior attacks on decoders, while remaining competitive with the strongest baselines in undefended settings. \method{} also scales to significantly longer sequences on encoder-only models, where prior attacks perform poorly or are computationally infeasible.

\method{} builds on the observation that gradients of transformer linear layers contain exploitable low-rank structure. In particular, for a linear layer, the row span of the layer inputs coincides with the column space of the corresponding weight gradient under mild rank conditions. Prior work uses this fact to test whether discrete token candidates lie in the gradient-induced subspace~\cite{dager}. In contrast, \method{} turns this signal into a differentiable objective, directly optimizing continuous token embeddings so that their hidden representations align with the subspaces revealed by the observed gradients. This continuous formulation makes the attack more robust to numerical imprecision and better suited for attacking challenging settings. The span-distance objective underlying this optimization is illustrated in panel A of \cref{fig:overview}.

This subspace view enables efficient recovery for both decoder- and encoder-based transformers. For decoder-only transformers, shown in panel B in \cref{fig:overview}, \method{} exploits causal attention: because the representation at a position depends only on preceding tokens, the attack can recover embeddings sequentially using subspace constraints from the first few transformer layers. For encoder-only transformers, shown in panel C in \cref{fig:overview}, causal recovery is unavailable. \method{} therefore introduces a bidirectional alignment objective, which not only aligns optimized embeddings to the observed gradient column space, but also incentivizes the observed gradient directions to be covered by the optimized embedding span. This bidirectional alignment discourages degenerate collapse, including repeated-token and duplicate-sequence solutions, and provides a stronger signal in non-causal architectures. 

By optimizing against gradient-induced subspaces rather than matching full model gradients, \method{} requires only partial forward and backward passes through the relevant layers, making reconstruction substantially more efficient. More importantly, the continuous objective avoids brittle discrete decisions and improves reconstruction in challenging regimes, including larger client batches, quantized gradients, and DP-style noise.

\subsection{Evaluation and Results} 

We evaluate \method{} on modern transformer language models, including \gemma{}~\cite{gemma3} and \embgemma{}~\cite{embeddinggemma}, covering both decoder-only and encoder-only architectures used in contemporary fine-tuning applications~\cite{medgemma,medgemma15,financialgemma,embeddinggemmaft}. On WikiText~\cite{wikitext} batches with varying batch sizes, sequence lengths, and defense levels, we measure reconstruction quality with ROUGE-1 and ROUGE-L~\cite{rouge}. Our evaluation focuses on the regimes where prior text inversion attacks are most brittle: batched updates, dense attention, and perturbed or lower-precision gradients.

In decoder-only models, \method{} is competitive with, and slightly outperforms, \dager{}~\cite{dager} in undefended settings, where \dager{} remains a strong baseline. However, under DP-style gradient noise, \dager{} collapses to near-zero reconstruction quality, while \method{} continues to recover meaningful text, achieving ROUGE-1 of more than $70\%$ up to $\sigma=10^{-3}$. On encoder-only models, where causal token-by-token recovery is infeasible, \method{} provides a larger improvement at a ROUGE-1 of $60\%$ on as many as 16-sequence batches. Finally, \method{} performs well even under quantization and a wide range of batch size--sequence length combinations.

\subsection{Key contributions} \noindent Our main contributions include:
\begin{itemize}
    \item \method{}, a robust and scalable continuous gradient inversion attack for transformer LLMs that recovers private text across a range of challenging settings.
    \item A differentiable embedding-subspace objective derived from the low-rank structure of transformer linear-layer gradients. The objective aligns optimized token embeddings with gradient-induced subspaces and extends to encoder-only models through a bidirectional alignment loss (\cref{sec:methodology}).
    \item An extensive evaluation on decoder- and encoder-only transformer models, including \gemma{} and \embgemma{}, showing that \method{} recovers substantially more text than prior gradient inversion attacks in defended and encoder-only settings, while remaining competitive with \dager{} in the undefended decoder regime, revealing further privacy vulnerabilities (\cref{sec:results}).
     \item An open-source implementation\footnote{\href{https://anonymous.4open.science/r/cont-dager-B6BD/README.md}{https://anonymous.4open.science/r/cont-dager-B6BD/README.md}} of \method{} to facilitate future research on privacy risks and defenses for collaborative LLM training.
\end{itemize}

%% file: paper_files/related.tex
\section{Related Work} \label{sec:related}

In this section, we review the work most relevant to our setting, including standard settings and approaches for gradient leakage (\cref{sec:related:gradient-inversion}, \cref{sec:related:optimization-inversion}), text-specific reconstruction methods (\cref{sec:related:text-recovery}), and low-rank inversion techniques (\cref{sec:related:exact-inversion}).

\subsection{Gradient Inversion} \label{sec:related:gradient-inversion}

Federated learning~\cite{federated} enables multiple clients to train a shared model through a central server without directly transmitting raw training data. However, this paradigm does not guarantee privacy, as the gradients or parameter updates sent by clients can encode substantial information about the examples used to compute them through gradient inversion attacks. These attacks are commonly distinguished by the adversary's control over the training process. In the honest-but-curious setting, the server follows the prescribed protocol but attempts to infer client data from the updates it observes \cite{dlg,idlg,rgap,towardsgeneral,geiping,tag,lamp,film,poolerattack,dager,spear,spear++}. In the malicious-server setting, the server actively modifies the model shared with the clients to amplify leakage \cite{robbingthefed,fishing,decepticons,panningforgold,minegrad}. Malicious attacks can enable substantially stronger extraction, but they rely on a stronger adversary and may be detectable when clients can audit the model or protocol. In this work, we focus on the more restrictive and realistic honest-but-curious setting.

\subsection{Optimization-based Gradient Inversion} \label{sec:related:optimization-inversion}
A central class of honest-but-curious attacks formulates reconstruction as an optimization problem. These methods initialize dummy inputs, and sometimes dummy labels, then iteratively update them so that the gradients they induce match the observed client gradients~\cite{dlg,geiping,idlg}. This approach has been especially successful in continuous domains such as images, where the input representation can be directly optimized and where image priors can regularize the reconstruction. 

Textual data poses additional challenges, as the input data is discrete, and as a result, optimized embeddings must be mapped to valid tokens in a model's vocabulary. The search space over discrete tokens also grows with vocabulary size, sequence length, batch size, and model hidden dimension. Consequently, direct optimization-based attacks often scale poorly to large language models and larger batches~\cite{tag,lamp,dager}.

\subsection{Recovering Textual Training Data} \label{sec:related:text-recovery}

Several works have adapted gradient inversion to the text domain by incorporating language-specific structure. TAG~\cite{tag} extends gradient-matching techniques to textual inputs, while later methods combine continuous optimization with language priors, discrete search, or model-specific signals to improve token recovery~\cite{lamp,film,poolerattack}. Other attacks exploit representations exposed by particular architectures or task heads, such as intermediate activations or pooling layers~\cite{poolerattack}.

\subsection{Exact and Low-Rank Gradient Inversion} \label{sec:related:exact-inversion}

A complementary line of work seeks exact or nearly exact recovery by exploiting algebraic structure in gradients. In the malicious setting, the server can design the model or training procedure so that client gradients encode private information more directly~\cite{robbingthefed,fishing,decepticons,panningforgold,minegrad}. These attacks can be highly effective, including at larger batch sizes, but they assume active manipulation of the training process.

In the honest-but-curious setting, exact recovery is more constrained. APRIL~\cite{april} demonstrates exact recovery for transformer models, but only for batch size one. Other methods exploit the low-rank structure of linear-layer gradients~\cite{dager,grain,spear,spear++}. When the batch size is small relative to the layer dimensions, this structure can reveal subspaces associated with individual training examples and can sometimes enable exact or near-exact batch recovery. However, these methods often rely on additional assumptions. SPEAR and SPEAR++ use architectural properties such as ReLU-induced sparsity, which limits their direct applicability to standard transformer models~\cite{spear,spear++}. DAGER~\cite{dager} extends low-rank inversion to transformer-based language models by exploiting token discreteness, but its recovery procedure depends on numerically sensitive token filters. While this makes the method more reliable with high-precision gradients, it can be brittle under lower quantization or DP-SGD~\cite{dpsgd} noise.

\method{} builds on the observation that linear-layer gradients contain exploitable low-rank structure, but uses this structure to guide a continuous optimization procedure rather than relying solely on exact algebraic filtering. This allows the attack to retain information from the low-rank signal while improving robustness across larger batches and perturbed gradients.




%% file: paper_files/background.tex
\section{Background and Notation}
\label{sec:background}

We consider the honest-but-curious federated learning setting, where a client computes gradients of a transformer model on a private batch of text and shares these gradients with a central server. The server follows the FL protocol but attempts to reconstruct the client's input from the observed update. In the rest of the section, we first recall how transformer models work (\cref{sec:background:transformers}), and then how their gradient structure is exploited by DAGER~\cite{dager}, which we also leverage in \method{} (\cref{sec:background:gradients}). For convenience, we include a summary of the notation used throughout the paper in \cref{tab:notation}.

\input{paper_files/tables/notation.tex}

\subsection{Transformers}
\label{sec:background:transformers}

We study transformer models with hidden dimension $\HiddenDim$, vocabulary $\Vocab$ of size $\VocabSize$, associated embedding vectors $\Embeddings=\{\TokenEmbedding{v}:v\in\Vocab\}$, and $\NumBlocks$ transformer blocks. A client batch contains $\BatchSize$ token sequences, of which the $b$-th sequence has length $\SeqLength{b}$. We write $\Token{b,i}\in\Vocab$ for the token at position $i$ of the $b$-th client sequence. After reconstruction, $\RecoveredToken{b,i}$ denotes the recovered token at the same position. We denote the length of the longest sequence by
\begin{equation}
	\MaxSeqLength = \max_{b \in \{1,\ldots,\BatchSize\}} \SeqLength{b},
\end{equation}
and the total number of non-padding tokens in the batch by
\begin{equation}
	\NumTokens = \sum_{b=1}^{\BatchSize} \SeqLength{b} .
\end{equation}
At each transformer block, the transformer architecture takes the embeddings of all tokens in a sequence as inputs and produces a new output embedding for each token using self-attention. We represent with $\HiddenState{l} \in \mathbb{R}^{\NumTokens \times \HiddenDim}$ the matrix of token embeddings input to the $l$-th transformer block, where each row corresponds to one token in the flattened client batch. The query, key, and value representations of $\HiddenState{l}$ in block $l$ are then given by:
\begin{equation}
	\QueryActivation{l} = \HiddenState{l} \QueryWeight{l},
	\qquad
	\KeyActivation{l} = \HiddenState{l} \KeyWeight{l},
	\qquad
	\ValueActivation{l} = \HiddenState{l} \ValueWeight{l},
	\label{eq:qkv}
\end{equation}
where $\QueryWeight{l}, \KeyWeight{l}, \ValueWeight{l} \in \mathbb{R}^{\HiddenDim \times \HiddenDim}$ are model parameters. Given an attention mask $\AttentionMask \in \mathbb{R}^{\NumTokens \times \NumTokens}$, the corresponding self-attention output is given by:
\begin{equation}
	\operatorname{Attn}(\QueryActivation{l},\KeyActivation{l},\ValueActivation{l};\AttentionMask)
	=
	\operatorname{softmax}
	\left(
	\frac{\QueryActivation{l} (\KeyActivation{l})^\top}{\sqrt{\HiddenDim}} + \AttentionMask
	\right)\ValueActivation{l},
	\label{eq:attention}
\end{equation}
where the softmax is applied row-wise. We use the additive-mask convention: entries of $\AttentionMask$ are $0$ for allowed attention edges and $-\infty$ for disallowed edges. The mask is block-diagonal across batch elements, preventing tokens from different client sequences from attending to one another.

For encoder-only models, $\AttentionMask$ is typically a padding mask, so each non-padding token may attend to all other non-padding tokens in the same sequence. Since encoder attention is non-causal, each representation generally depends on the full sequence. For decoder-only models, $\AttentionMask$ additionally contains a causal mask, ensuring that the representation at position $i$ can depend only on positions up to $i$.

We denote the full transformation computed by transformer block $l$ with $\TransformerBlock{l}: \mathbb{R}^{\NumTokens \times \HiddenDim} \times \mathbb{R}^{\NumTokens \times \NumTokens} \to \mathbb{R}^{\NumTokens \times \HiddenDim}$, where:
\begin{equation}
	\HiddenState{l+1} = \TransformerBlock{l}(\HiddenState{l};\AttentionMask),
	\qquad
	l \in \{1,\ldots,\NumBlocks\}.
	\label{eq:block-forward}
\end{equation}
In particular, $\TransformerBlock{l}$ includes the self-attention operation from Eq.~\ref{eq:attention}, as well as the standard model-specific components surrounding it, such as RoPE positional embeddings, multi-head projections, residual connections, LayerNorms, MLP layers, and any additional normalization or projection operations leading to the next block's projection. We abstract these components into $\TransformerBlock{l}$ since they are fixed, known parts of the model and can therefore be evaluated by the attacker during reconstruction.

We distinguish the raw input embedding matrix from the input to the first transformer block. Let $\HiddenState{0} \in \mathbb{R}^{\NumTokens \times \HiddenDim}$ denote the matrix of token embeddings before any model-specific preprocessing. For the true client batch, $\HiddenState{0}$ is obtained by applying the model's embedding lookup to the private token sequences, so each row is some vocabulary embedding $\TokenEmbedding{v}\in\Embeddings$ for a token $v\in\Vocab$. In \method{}, the attacker instead introduces a continuous variable $\OptimizedInput{0} \in \mathbb{R}^{\NumTokens \times \HiddenDim}$ and optimizes it to recover the client's input embeddings.

We denote by $\Preprocess$ the known preprocessing applied before the first transformer block, such as adding absolute positional embeddings or applying any normalization. Thus,
\begin{equation}
	\HiddenState{1} = \Preprocess(\HiddenState{0};\AttentionMask).
	\label{eq:embedding-preprocessing}
\end{equation}

Analogously, given $\OptimizedInput{0}$, we write $\OptimizedInput{l}$ for the recovered hidden states obtained by applying the same known preprocessing and transformer blocks to the optimized input, i.e., $\OptimizedInput{l+1}=\TransformerBlock{l}(\OptimizedInput{l};\AttentionMask)$. We write $\HiddenState{l}_i \in \mathbb{R}^{\HiddenDim}$ to denote the $i$-th row, i.e., token embedding, of $\HiddenState{l}$. When using batch-position indices, $\OptimizedInput{l}_{b,i}\in\mathbb{R}^{\HiddenDim}$ denotes the recovered hidden state at layer $l$ corresponding to the $i$-th token of the $b$-th client sequence.

\subsection{Self-Attention Gradients}
\label{sec:background:gradients}

The preceding notation identifies $\HiddenState{l}$ as the input to transformer block $l$, and in particular as the input to the query, key, and value projections in Eq.~\ref{eq:qkv}. Since the query projection in block $l$ is
\begin{equation}
	\QueryActivation{l} = \HiddenState{l} \QueryWeight{l},
\end{equation}
the gradient of a training loss $\TrainingLoss$ with respect to the query projection weights factors as:
\begin{equation}
	\GradQueryWeight{l}
	=
	(\HiddenState{l})^\top
	\GradQueryActivation{l},
	\label{eq:query-gradient-factorization}
\end{equation}
as shown by DAGER~\cite{dager}. Analogous identities hold for the key and value projections. Since $\HiddenState{l} \in \mathbb{R}^{\NumTokens \times \HiddenDim}$ and $\GradQueryActivation{l} \in \mathbb{R}^{\NumTokens \times \HiddenDim}$, Eq.~\ref{eq:query-gradient-factorization} implies that:
\begin{equation}
	\operatorname{rank}
	\left(
	\GradQueryWeight{l}
	\right)
	\leq \NumTokens .
\end{equation}
Thus, DAGER~\cite{dager} shows that whenever the total number of non-padding tokens satisfies $\NumTokens < \HiddenDim$, the query-gradient matrix is rank-deficient. DAGER~\cite{dager} further shows that, under a mild full-rank assumption on the backpropagated query gradients, this low-rank gradient reveals the span of the block inputs $\HiddenState{l}$:

\begin{theorem}[Theorem 5.1 of DAGER~\cite{dager}]\label{theorem:dager}
	If $\NumTokens < \HiddenDim$ and the matrix $\GradQueryActivation{l}$ is of full rank $\NumTokens$, then
	\begin{equation}
		\operatorname{rowspan}(\HiddenState{l})
		=
		\operatorname{colspan}
		\left(
		\GradQueryWeight{l}
		\right).
	\end{equation}
\end{theorem}
This theorem implies that every row of $\HiddenState{l}$ lies in the column span of the observed query-gradient matrix. We denote this gradient-induced span, an orthonormal basis for it, and the projection matrix it induces by
\begin{equation}
	\GradientSubspace{l}
	=
	\operatorname{colspan}
	\left(
	\GradQueryWeight{l}
	\right),
	\,
	\SubspaceBasis{l}
	=
	\operatorname{basis}(\GradientSubspace{l}),
	\,
	\SubspaceProjection{l}
	=
	\operatorname{proj}\GradientSubspace{l} .
	\label{eq:gradient-span}
\end{equation}
In particular, we have $\SubspaceProjection{l}=(\SubspaceBasis{l})^\top \SubspaceBasis{l}$.

DAGER exploits \cref{theorem:dager} by testing whether discrete candidate token or sequence embeddings belong to this gradient-induced subspace $\GradientSubspace{l}$. Since the subspace has dimension at most $\NumTokens$, incorrect candidates are unlikely to lie exactly in it when $\NumTokens < \HiddenDim$.

In finite precision, DAGER replaces exact span membership with a distance-to-subspace computation. For a candidate embedding vector $z \in \mathbb{R}^{\HiddenDim}$, DAGER defines the following distance (Equation~2 in \cite{dager}):
\begin{equation}
	\SpanDistance(z,\SubspaceProjection{l})
	=
	\left\|
	z - 
	 z (\SubspaceProjection{l})^\top
	\right\|_2 .
	\label{eq:dager-distance}
\end{equation}

In practice, the projection is computed using an orthonormal basis for the dominant singular subspace of $\GradQueryWeight{l}$, after truncating singular values below a numerical threshold. DAGER then treats a candidate as consistent with the observed gradient when $\SpanDistance(z,\SubspaceProjection{l})$ is below a chosen layer-dependent tolerance.

\method{} uses the same distance signal in Eq.~\ref{eq:dager-distance}, but avoids turning it into a hard discrete membership test. Instead of enumerating candidate tokens or sequences and checking whether their embeddings pass a threshold, \method{} directly optimizes the input embedding matrix $\OptimizedInput{0}$ so that the representations induced by the transformer forward maps approach the gradient-induced subspaces. The continuous reconstruction objective built from Eq.~\ref{eq:dager-distance} is introduced in the next section. In \cref{tab:notation}, we further provide a succinct summary of all notations used throughout the paper.

\subsection{Threat Model}

We follow the standard assumptions in the honest-but-curious federated learning setting, where the attacker knows the model architecture and current parameter values, and observes the client's gradients for all trainable parameters except the embedding layer. Following prior optimization-based gradient-inversion attacks~\cite{lamp,tag,dlg,geiping}, we also assume that the attacker knows the input shape, namely the batch size $\BatchSize$ and sequence lengths $\{\SeqLength{b}\}_{b=1}^{\BatchSize}$. These quantities determine the total number of non-padding tokens $\NumTokens$ and the attention mask $\AttentionMask$ for both encoder and decoder models. Unlike most optimization-based attacks, \method{} does not require the true training labels, nor does it optimize over dummy labels during reconstruction.

%% file: paper_files/tables/notation.tex
\begin{table*}[!t]
	\centering
	\newcommand{\twocol}[1]{\begin{tabular}[c]{@{}l@{}}#1\end{tabular}}
	\resizebox{1.0\linewidth}{!}{
		\begin{tabular}{cl@{\hskip 1cm}cl}
			\toprule
			\textbf{Symbol} & \multicolumn{1}{l}{\textbf{Definition}} &
			\textbf{Symbol} & \multicolumn{1}{l}{\textbf{Definition}} \\
			\midrule
			
			$\BatchSize$ &
			Batch size &
			$\NumBlocks$ &
			Number of transformer blocks \\
			
			$\Vocab$ &
			Vocabulary set &
			$\VocabSize$ &
			Vocabulary size, $\VocabSize = |\Vocab|$ \\
			
			$\Embeddings$ &
			Embedding vectors of the tokens in $\Vocab$ &
			$\TokenEmbedding{v}$ &
			The embedding vector associated with token $v\in\Vocab$ \\
			
			$\TrainingLoss$ &
			Training loss function used by clients &
			$\SeqLength{b}$ &
			Token length for the $b$-th sequence \\
			
			$\MaxSeqLength$ &
			\twocol{Length of the longest sequence, $\MaxSeqLength=\max_b \SeqLength{b}$} &
			$\NumTokens$ &
			\twocol{Total number of non-padding tokens, $\NumTokens=\sum_{b=1}^{\BatchSize} \SeqLength{b}$} \\
			
			$\HiddenState{0}$ &
			Raw transformer input embeddings &
			$\HiddenState{l}$ &
			Hidden states input to the $l$-th attention layer \\
			
			$\HiddenDim$ &
			Transformer hidden dimension &
			$\AttentionMask$ &
			Attention mask \\
			
			$\QueryWeight{l}, \KeyWeight{l}, \ValueWeight{l}$ &
			\twocol{Query/Key/Value projection matrices\\for the $l$-th attention layer} &
			$\QueryActivation{l}, \KeyActivation{l}, \ValueActivation{l}$ &
			\twocol{The Query/Key/Value activations\\in the $l$-th attention layer} \\
			
			$\Preprocess$ &
			\twocol{Preprocessing done on the raw embeddings\\before feeding them into the first attention block} &
			$\TransformerBlock{l}$ &
			\twocol{$l$-th transformer block, i.e., $\HiddenState{l+1}=\TransformerBlock{l}(\HiddenState{l};\AttentionMask)$} \\
			
			$\NumInitializations$ &
			Number of random initializations used in \method{} &
			$\NumSteps$ &
			\method{}'s number of continuous optimization steps \\
			
			$\LearningRate$ &
			\method{}'s learning rate &
			$\OptimizedInput{0}$ &
			Continuous raw input embeddings optimized by \method{} \\
			
			$\OptimizedInput{l}$ &
			\twocol{Recovered hidden states obtained from $\OptimizedInput{0}$,\\ $\OptimizedInput{l}=\TransformerBlock{l-1}(\OptimizedInput{l-1};\AttentionMask)$} &
			$\OptimizedInput{l}_{b,i}$ & 
			\twocol{Recovered hidden state at layer $l$\\corresponding to $i$-th token in the $b$-th sequence}\\
			
			$\GradientSubspace{l}$ &
			Gradient subspace at layer $l$, $\GradientSubspace{l}=\operatorname{colspan}\left(\GradQueryWeight{l}\right)$ &
			$\RecoveredSubspace{l}$ &
			\twocol{Recovered activation subspace at layer $l$,\\
				$\RecoveredSubspace{l}=\operatorname{rowspan}(\OptimizedInput{l})$} \\

			$\SubspaceBasis{l}$ & 
			Orthonormal basis of $\GradientSubspace{l}$ &
			$(\InitMu{i},\InitSigma{i})$ &
			Gaussian prior for tokens at position $i$\\
			
			$\SubspaceProjection{l}$ &
			Projection matrix onto $\GradientSubspace{l}$ &
			$\RecoveredSubspaceProjection{l}$ &
			Projection matrix onto $\RecoveredSubspace{l}$ \\
			
			$\SubspaceDedupProjection{l,b}$ &
			Projection matrix onto $\GradientSubspace{l} \cap \OptimizedInput{l}_{1:b-1,1}$ &
			$\Loss$ &
			\method{}'s final loss\\
			
			$\ForwardLoss$ &
			Forward span-distance loss used by \method{} &
			$\BackwardLoss$ &
			Backward span-distance loss used by \method{} \\
			
			$\DedupLoss$ &
			\twocol{Loss term penalizing repeating tokens\\ during decoder first-token recovery} & 
			$\DedupWeightCoeff$ &
			\twocol{Weight of the $\DedupLoss$ loss term used during first-token\\ recovery for decoder models} \\
			
			$\InvJitterCoeff$ &
			\twocol{Small diagonal jitter used to stabilize matrix\\inversions in $\BackwardLoss$} &
			$\BackwardCoeff$ &
			\twocol{Weight of the $\BackwardLoss$ loss term used for encoder models} \\
			
			$\SpanDistance(z,\SubspaceProjection{})$ &
			Distance from the vector $z$ to its projection using $\SubspaceProjection{}$&
			$\MaxLayers$ &
			Number of transformer layers used by \method{}\\

			$\Token{b,i}$ & 
			Token for $i$-th position of the $b$-th client sequence &
			$\RecoveredToken{b,i}$ &
			Recovered token for $i$-th position of the $b$-th client sequence \\
			
			\bottomrule
		\end{tabular}
	}
	\caption{Table of notations used in the technical description of \method{}.}
	\label{tab:notation}
\end{table*}

%% file: paper_files/methodology.tex
\section{\method{}} \label{sec:methodology}

We now detail the \method{} attack procedure. We first describe the shared components across encoder and decoder attacks, including the recovery objective (\cref{sec:methodology:objective}) and the initialization strategy (\cref{sec:methodology:initialization}). We then give the specific algorithms for the decoder and encoder attacks in \cref{sec:methodology:decoder} and \cref{sec:methodology:encoder}, respectively.

\subsection{Recovery Objective} \label{sec:methodology:objective}

The common structure of \method{} is the same for both encoders and decoders. For each attacked layer $l\leq \MaxLayers$, we use the gradient-induced subspace basis $\SubspaceBasis{l}$ obtained from the observed gradients $\GradQueryWeight{l}$ via SVD as described in \cref{theorem:dager} and the original DAGER paper~\cite{dager}. We assume $\SubspaceBasis{l}$ stores an orthonormal basis of this subspace as rows, and therefore the corresponding projection matrix is given by
$
\SubspaceProjection{l}=(\SubspaceBasis{l})^\top\SubspaceBasis{l}.
$
Since the input shape is assumed known, the attacker also knows $\NumTokens$, which determines the rank of the SVD used to extract $\SubspaceBasis{l}$.

In \method{}, we optimize raw input embeddings $\OptimizedInput{0}$ so that their respective hidden states $\OptimizedInput{l}$ induced by the known transformer blocks $\TransformerBlock{l}$ align with the observed gradient-induced subspaces for all $l\leq \MaxLayers$. 
Since the subspace constraint from \cref{theorem:dager} only identifies vector directions rather than their norms, we normalize each recovered hidden state vector $\OptimizedInput{l}_{b,i}$ before measuring its distance $\SpanDistance$ to its projection with $\SubspaceProjection{l}$. This normalization also prevents the trivial unnormalized solution in which the optimizer drives hidden-state norms toward $0$, thereby reducing the distance without recovering a meaningful direction. Our forward span-distance loss is then simply given by
\begin{equation}
	\ForwardLoss
	=
	\sum_{l=1}^{\MaxLayers}
	\sum_{(b,i)\in\Omega}
	\SpanDistance
	\left(
	\frac{\OptimizedInput{l}_{b,i}}{\|\OptimizedInput{l}_{b,i}\|_2},
	\SubspaceProjection{l}
	\right)^2,
	\label{eq:method-forward-loss}
\end{equation}
where $\Omega$ denotes the token positions that are active in the current optimization instance. For encoder attacks, $\Omega=\{(b,i):1\leq b\leq \BatchSize,\ 1\leq i\leq \SeqLength{b}\}$, so the full batch is optimized jointly. For decoder attacks at recovery step $(b,i)$, $\Omega=\{(b,i)\}$, since due to causal masks, similarly to DAGER~\cite{dager}, we can attack individual tokens, provided our attack has already recovered all preceding tokens in the sequence. In $\ForwardLoss$, we sum over multiple layers to provide additional constraints and stabilize our optimization, while using only the first $\MaxLayers$ layers to keep the attack computationally efficient.

After the continuous optimization ends, we map each recovered embedding to the nearest token embedding vector via cosine similarity, as described in \cref{alg:project-to-vocab}.

\begin{algorithm}[t]
	\caption{\textsc{ProjectToVocab}: Cosine Projection onto Token Embeddings}
	\label{alg:project-to-vocab}
	\begin{algorithmic}[1]
		\REQUIRE Recovered continuous embeddings $\RecoveredInput{0}$,\\
		\quad\, Sequence lengths $\{\SeqLength{b}\}_{b=1}^{\BatchSize}$, \\
		\quad\, Vocabulary embeddings $\{\TokenEmbedding{v}\}_{v\in\Vocab}$
		\ENSURE Reconstructed token sequences $\{\hat{t}_{b,i}\}$
		
		\FOR{$b = 1$ to $\BatchSize$}
		\FOR{$i = 1$ to $\SeqLength{b}$}
		
		\STATE $
		\RecoveredToken{b,i}
		\leftarrow
		\underset{v\in\Vocab}{\operatorname{argmax}}
		\operatorname{CosSim}
		\left(
		\RecoveredInput{0}_{b,i},
		\TokenEmbedding{v}
		\right)
		$
		
		\ENDFOR
		\ENDFOR
		
		\RETURN $\{\RecoveredToken{b,i}\}$
		
	\end{algorithmic}
\end{algorithm}

\subsection{Initialization} \label{sec:methodology:initialization}

The objective in \cref{eq:method-forward-loss} is highly non-convex, and, thus, the objective value to which it converges is highly dependent on the initialization of the optimization variable $\OptimizedInput{0}$. As we are aiming to arrive at the global minimum of \cref{eq:method-forward-loss}, which is 0 in undefended scenarios, \method{} does $\NumInitializations$ reconstructions from $\NumInitializations$ independent initializations. Additionally, we use a lightweight data prior to ensure we initialize $\OptimizedInput{0}$ only to promising values. In particular, we build fixed-length tokenized windows from a large corpus such as \wikitext{} and, for each absolute token position $i \in \{1,\ldots,\MaxSeqLength\}$, fit a full-covariance Gaussian $\mathcal{N}(\InitMu{i},\InitSigma{i})$ to the raw embedding vectors observed at that position across text sequences in the dataset. Each independent initialization samples its raw initial vectors $\OptimizedInput{0}_{b,i}$ from $\mathcal{N}(\InitMu{i},\InitSigma{i})$. We use this prior as the default initializer throughout the paper, and compare it with alternative position-agnostic initializers in \cref{sec:results:ablations:init}.

\subsection{Decoder Attack} \label{sec:methodology:decoder}

For decoder-only transformers, we leverage the causal attention mask to recover tokens sequentially. Under the decoder mask $\AttentionMask_{\mathrm{dec}}$, the hidden states of a prefix are independent of tokens that appear later in the same sequence. Formally, for any prefix length $i$ and any layer $l$,
\begin{equation}
	\TransformerBlock{l}
	\left(
	\OptimizedInput{l}_{b,1:i};
	\AttentionMask_{\mathrm{dec}}
	\right)
	=
	\TransformerBlock{l}
	\left(
	\OptimizedInput{l};
	\AttentionMask_{\mathrm{dec}}
	\right)_{b,1:i}.
	\label{eq:decoder-prefix-property}
\end{equation}
We can therefore optimize the loss $\ForwardLoss$ for individual token embedding vectors, provided the respective sequence prefixes have been recovered in prior iterations of the attack. The loss for optimizing the embedding of token $i$ in sequence $b$, $\OptimizedInput{0}_{b,i}$, is given by
\begin{equation}
	\ForwardLoss^{(b,i)}
	=
	\sum_{l=1}^{\MaxLayers}
	\SpanDistance
	\left(
	\frac{\OptimizedInput{l}_{b,i}}{\|\OptimizedInput{l}_{b,i}\|_2},
	\SubspaceProjection{l}
	\right)^2,
	\label{eq:decoder-forward-loss}
\end{equation}
which sums the subspace span-distance losses across layers for the current position.

\subsubsection{First-token deduplication}

In the decoder setting, recovering the first token of each sequence is crucial because all subsequent token representations depend on it. If the first token of a sequence is not recovered correctly, the attack often fails on the rest of that sequence. Moreover, under the causal mask $\AttentionMask_\text{dec}$, first-token embeddings are independent across batch elements, so the forward span-distance loss $\ForwardLoss$ has multiple global minima corresponding to the ground-truth first-token embeddings $\HiddenState{0}_{:,1}$. To recover as many of these solutions as possible, we optimize the $\BatchSize$ client sequences one after another and add a deduplication loss $\DedupLoss$ that encourages the current first-token embedding $\RecoveredInput{0}_{b,1}$ to differ from previously recovered first-token embeddings $\RecoveredInput{0}_{1:b-1,1}$.

Suppose we are recovering the first token of sequence $b$. For each layer $1\leq l\leq \MaxLayers$, we use \textsc{DedupProjs} to construct a projection matrix $\SubspaceDedupProjection{l,b}$ onto the part of the gradient-induced subspace $\GradientSubspace{l}$ already occupied by the previously recovered first tokens $\RecoveredInput{0}_{1:b-1,1}$. As detailed in \cref{sec:dedup-projs}, this is done by forwarding $\RecoveredInput{0}_{1:b-1,1}$ through the attacked transformer layers $\{\TransformerBlock{l}\}^{\MaxLayers-1}_{l=0}$ and rotating the orthonormal basis $\SubspaceBasis{l}$ to align with the resulting hidden states $\RecoveredInput{l}_{1:b-1,1}$. The projection $\SubspaceDedupProjection{l,b}$ therefore identifies directions already used by earlier recovered first tokens.

We penalize the component of the current first-token hidden state $\OptimizedInput{l}_{b,1}$ along these directions. The deduplication loss for recovering the first token of sequence $b$ is
\begin{equation}
	\DedupLoss^{b}
	=
	\sum_{l=1}^{\MaxLayers}
	\left\|
	\frac{
		\OptimizedInput{l}_{b,1}
	}{
		\left\|
		\OptimizedInput{l}_{b,1}
		\right\|_2
	}
	(\SubspaceDedupProjection{l,b})^\top
	\right\|_2^2.
	\label{eq:decoder-dedup-loss}
\end{equation}
This loss is used only when recovering the first token of sequences after the first one. Thus, for $b>1$, the first-token decoder objective is
\begin{equation}
	\Loss^{(b,1)}
	=
	\ForwardLoss^{(b,1)}
	+
	\DedupWeightCoeff
	\DedupLoss^{b},
	\label{eq:decoder-first-token-loss}
\end{equation}
where $\DedupWeightCoeff$ controls the strength of the deduplication loss. For all other decoder recovery steps, namely positions $i>1$ and the first sequence $b=1$, we optimize only the forward span-distance loss:
\begin{equation}
	\Loss^{(b,i)}
	=
	\ForwardLoss^{(b,i)}.
	\label{eq:decoder-nondedup-loss}
\end{equation}
Additional implementation details and pseudocode for \textsc{DedupProjs} are given in \cref{sec:dedup-projs}.

\subsubsection{Algorithm description}

Having defined all of our objective components, we present our full decoder attack in \cref{alg:decoder-recover}. We recover the batch sequence-by-sequence, where we greedily recover each sequence token-by-token. For each token, we draw $\NumInitializations$ independent initial embeddings from the position-dependent prior  $\mathcal{N}(\InitMu{i},\InitSigma{i})$ (line 8). Each raw candidate embedding vector $z$ is then updated for $\NumSteps$ optimization steps based on the computed gradients $\nabla_{z}\Loss$ (line 18). In addition to the forward span-distance loss $\ForwardLoss$ computed from the induced hidden states $\OptimizedInput{l}$ (lines 10--14), the first token of each sequence also includes the deduplication loss $\DedupLoss$ when $b>1$ (lines 15--17). We keep the candidate $z^{\star}$ with the lowest final loss $\Loss^{\star}$ across all independent initializations (lines 19--21), and we use it as our recovered continuous embedding $\RecoveredInput{0}_{b,i}$ for that token. After all token embeddings are recovered, they are projected to the nearest vocabulary embeddings to obtain the final discrete reconstruction $\{\RecoveredToken{b,i}\}$ (line 23).

\begin{algorithm}[t]
	\caption{\textsc{DecoderRecovery}: Sequential Recovery for Decoder Models}
	\label{alg:decoder-recover}
	\begin{algorithmic}[1]
		\REQUIRE Transformer blocks $\{\TransformerBlock{l}\}^{\MaxLayers-1}_{l=0}$,\\
				 \quad\, Subspace projections $\{\SubspaceProjection{l}\}_{l=1}^{\MaxLayers}$, \\
				 \quad\, Subspace orthonormal basis $\{\SubspaceBasis{l}\}_{l=1}^{\MaxLayers}$, \\	
		         \quad\, Attention mask $\AttentionMask$, Sequence lengths $\{\SeqLength{b}\}_{b=1}^{\BatchSize}$, \\
		         \quad\, Vocabulary embeddings  $\{\TokenEmbedding{v}\}_{v\in\Vocab}$, \\
		         \quad\, Per-position Gaussian priors $\{(\InitMu{i}, \InitSigma{i})\}^{\MaxSeqLength}_{i=1}$
		\ENSURE Reconstructed token sequences $\{\RecoveredToken{b,i}\}$
		\STATE $\RecoveredInput{0}\leftarrow\bm{0}_{\NumTokens\times\HiddenDim}$ 
		\FOR{$b = 1$ to $\BatchSize$}
		
		\STATE $\{\SubspaceDedupProjection{l,b}\}_{l=1}^{\MaxLayers}
		\leftarrow
\textsc{DedupProjs}
		\left(
		\{\SubspaceBasis{l}\}_{l=1}^{\MaxLayers},
		\RecoveredInput{0}_{1:b-1,1},
		\AttentionMask
		\right)$
		
		\FOR{$i = 1$ to $\SeqLength{b}$}
		
		\STATE $\Loss^\star \leftarrow +\infty$
		\STATE $z^\star \leftarrow \emptyset$
		
		\FOR{$r = 1$ to $\NumInitializations$}
		
		\STATE $z \sim \mathcal{N}(\InitMu{i},\InitSigma{i})$
		
		\FOR{$s = 1$ to $\NumSteps$}
		
		\STATE
		$
		\OptimizedInput{0}
		\leftarrow
		\left[
		\RecoveredInput{0}_{b,1:i-1}
		\,;\,
		z
		\right]
		$
		
		\FOR{$l = 1$ to $\MaxLayers$}
			\STATE$
			\OptimizedInput{l}
			\leftarrow
			\TransformerBlock{l-1}
			\left(
			\OptimizedInput{l-1};
			\AttentionMask
			\right)
			$
		\ENDFOR
		\STATE $
		\ForwardLoss
		\leftarrow
		\sum_{l=1}^{\MaxLayers}
		\SpanDistance
		\left(
		\frac{
			\OptimizedInput{l}_{b,i}
		}{
			\left\|
			\OptimizedInput{l}_{b,i}
			\right\|_2
		},
		\SubspaceProjection{l}
		\right)^2
		$
		\STATE$
		\Loss
		\leftarrow
		\ForwardLoss
		$
		\IF{$i == 1$ and $b > 1$}
			\STATE$
			\DedupLoss
			\leftarrow
			\sum_{l=1}^{\MaxLayers}
			\left\|
			\frac{
				\OptimizedInput{l}_{b,1}
			}{
				\left\|
				\OptimizedInput{l}_{b,1}
				\right\|_2
			}
			(\SubspaceDedupProjection{l,b})^\top
			\right\|^2_2
			$
			\STATE$
			\Loss
			\leftarrow
			\Loss
			+
			\DedupWeightCoeff
			\DedupLoss
			$
		\ENDIF

		\STATE $z\leftarrow z - \LearningRate\nabla_{z}\Loss$
		
		\ENDFOR
		
		\IF{$\Loss < \Loss^\star$}
			\STATE $\Loss^\star \leftarrow \Loss$
			\STATE $z^\star \leftarrow z$
		\ENDIF
		
		\ENDFOR
		
		\STATE $\RecoveredInput{0}_{b,i} \leftarrow z^\star$
		
		\ENDFOR
		
		\ENDFOR
		
		\RETURN $\textsc{ProjectToVocab}(\RecoveredInput{0},\{\SeqLength{b}\}_{b=1}^{\BatchSize},\{\TokenEmbedding{v}\}_{v\in\Vocab})$
		
	\end{algorithmic}
\end{algorithm}

\subsection{Encoder Attack} \label{sec:methodology:encoder}

Encoder-only models use dense, non-causal self-attention, so all hidden states within a sequence are interdependent. Changing one token can change the hidden states of all other tokens in the same sequence. Unlike the decoder case, we cannot use a causal ordering to recover text sequences one token at a time. We therefore jointly optimize the entire raw token embedding matrix $\OptimizedInput{0}$.

\subsubsection{Backward span-distance loss}
For larger batch sizes $\BatchSize>1$, using only the forward span-distance loss $\ForwardLoss$ from \cref{eq:method-forward-loss} can lead to repeated or degenerate solutions. The reason is that $\ForwardLoss$ does not require the recovered hidden states to cover the whole observed subspace $\GradientSubspace{l}$. Instead, several recovered tokens or even whole sequences can collapse to similar directions, while still incurring a small forward loss. To mitigate this, we introduce a backward span-distance loss $\BackwardLoss$ that enforces the reverse inclusion: the gradient-induced basis directions $\SubspaceBasis{l}$ should also lie in the span of the recovered hidden states $\RecoveredSubspace{l}$. This encourages the recovered batch to cover the full observed subspaces $\GradientSubspace{l}$ and penalizes low-rank repeated-token solutions.

Using the current recovered hidden states $\OptimizedInput{l}$, we first derive a differentiable, regularized projection matrix onto the span of the optimized hidden states
\begin{equation}
\RecoveredSubspaceProjection{l}
=
\bigl(\OptimizedInput{l}\bigr)^\top
\left(
\OptimizedInput{l}\bigl(\OptimizedInput{l}\bigr)^\top + \InvJitterCoeff I
\right)^{-1}
\OptimizedInput{l},
\label{eq:encoder-recovered-projection}
\end{equation}
where the coefficient $\InvJitterCoeff$ is a small diagonal jitter used to avoid rank collapse during optimization. We note that we avoid explicitly generating the orthonormal basis $\SubspaceBasis{l}$ here, and instead we generate $\RecoveredSubspaceProjection{l}$ directly, as in our experiments differentiating through the SVD required to construct $\SubspaceBasis{l}$ significantly slowed down our attack. Based on $\RecoveredSubspaceProjection{l}$, we define the backward span-distance loss:
\begin{equation}
\BackwardLoss
=
\sum_{l=1}^{\MaxLayers}
\sum_{i=1}^{\NumTokens}
\SpanDistance
\left(
\SubspaceBasis{l}_{i},
\RecoveredSubspaceProjection{l}
\right)^2.
\label{eq:encoder-backward-loss}
\end{equation}
The inner sum ranges over the orthonormal basis vectors, i.e., the rows of $\SubspaceBasis{l}$. Since these basis vectors are already orthonormal, no additional normalization is required. The full encoder loss is a weighted sum of the forward and backward losses:
\begin{equation}
\Loss
=
\ForwardLoss
+
\BackwardCoeff\,\BackwardLoss.
\label{eq:encoder-total-loss}
\end{equation}

\subsubsection{Algorithm description}
Algorithm~\ref{alg:encoder-recover} optimizes $\Loss$ directly. As described in \cref{sec:methodology:initialization}, we run $\NumInitializations$ independent optimizations. For each independent initialization, the full batch of embeddings $\OptimizedInput{0}$ is sampled from the per-position Gaussian priors $\{(\InitMu{i}, \InitSigma{i})\}^{\MaxSeqLength}_{i=1}$ (lines 4--7), with the optimized embedding matrix $\OptimizedInput{0}$ then forwarded jointly through the attacked transformer blocks $\{\TransformerBlock{l}\}^{\MaxLayers-1}_{l=0}$ (lines 9--10). The forward and backward losses are then computed from the resulting hidden states $\OptimizedInput{l}$ (lines 11--14), followed by updating the full matrix $\OptimizedInput{0}$ based on the computed gradients $\nabla_{\OptimizedInput{0}}\Loss$ (line 15). The final continuous solution $\RecoveredInput{0}$ is selected based on the final encoder loss $\Loss$ (lines 16--18) and is projected back to tokens $\{\RecoveredToken{b,i}\}$ only after the joint optimization has converged (line 19).
\begin{algorithm}[t]
	\caption{\textsc{EncoderRecovery}: Joint Embedding Recovery for Encoder Models}
	\label{alg:encoder-recover}
	\begin{algorithmic}[1]
		\REQUIRE Transformer blocks $\{\TransformerBlock{l}\}^{\MaxLayers-1}_{l=0}$,\\
		\quad\, Subspace projections $\{\SubspaceProjection{l}\}_{l=1}^{\MaxLayers}$, \\
		\quad\, Subspace orthonormal basis $\{\SubspaceBasis{l}\}_{l=1}^{\MaxLayers}$, \\		
		\quad\, Attention mask $\AttentionMask$, Sequence lengths $\{\SeqLength{b}\}_{b=1}^{\BatchSize}$, \\
		\quad\, Vocabulary embeddings $\{\TokenEmbedding{v}\}_{v\in\Vocab}$, \\
		\quad\, Per-position Gaussian priors $\{(\InitMu{i}, \InitSigma{i})\}^{\MaxSeqLength}_{i=1}$
		\ENSURE Reconstructed token sequences $\{\RecoveredToken{b,i}\}$
		
		\STATE $\Loss^\star \leftarrow +\infty$
		\STATE $\RecoveredInput{0}\leftarrow\bm{0}_{\NumTokens\times\HiddenDim}$
		
		\FOR{$r = 1$ to $\NumInitializations$}
		
		\STATE $\OptimizedInput{0}\leftarrow\bm{0}_{\NumTokens\times\HiddenDim}$
		
		\FOR{$b = 1$ to $\BatchSize$}
		\FOR{$i = 1$ to $\SeqLength{b}$}
		\STATE $
		\OptimizedInput{0}_{b,i}
		\sim
		\mathcal{N}(\InitMu{i},\InitSigma{i})
		$
		\ENDFOR
		\ENDFOR
		
		\FOR{$s = 1$ to $\NumSteps$}
		
		\FOR{$l = 1$ to $\MaxLayers$}
		\STATE$
		\OptimizedInput{l}
		\leftarrow
		\TransformerBlock{l-1}
		\left(
		\OptimizedInput{l-1};
		\AttentionMask
		\right)
		$
		\STATE $
		\RecoveredSubspaceProjection{l}\leftarrow(\OptimizedInput{l})^\top (\OptimizedInput{l} (\OptimizedInput{l})^\top + \InvJitterCoeff I )^{-1} \OptimizedInput{l} 
		$
		\ENDFOR
		
		\STATE $
		\ForwardLoss
		\leftarrow
		\sum_{l=1}^{\MaxLayers}
		\sum_{b=1}^{\BatchSize}
		\sum_{i=1}^{\SeqLength{b}}
		\SpanDistance
		\left(
		\frac{
			\OptimizedInput{l}_{b,i}
		}{
			\left\|
			\OptimizedInput{l}_{b,i}
			\right\|_2
		},
		\SubspaceProjection{l}
		\right)^2
		$		
		\STATE $
		\BackwardLoss
		\leftarrow
		\sum_{l=1}^{\MaxLayers}
		\sum_{i=1}^{\NumTokens}
		\SpanDistance
		\left(
		\SubspaceBasis{l}_{i},
		\RecoveredSubspaceProjection{l}
		\right)^2
		$
		
		\STATE $
		\Loss
		\leftarrow
		\ForwardLoss
		+
		\BackwardCoeff
		\BackwardLoss
		$
		
		\STATE $\OptimizedInput{0}\leftarrow \OptimizedInput{0} - \LearningRate\nabla_{\OptimizedInput{0}}\Loss$
		
		\ENDFOR
		
		\IF{$\Loss < \Loss^\star$}
		\STATE $\Loss^\star \leftarrow \Loss$
		\STATE $\RecoveredInput{0}\leftarrow \OptimizedInput{0}$
		\ENDIF
		
		\ENDFOR
		
		\RETURN $\textsc{ProjectToVocab}(\RecoveredInput{0},\{\SeqLength{b}\}_{b=1}^{\BatchSize},\{\TokenEmbedding{v}\}_{v\in\Vocab})$
		
	\end{algorithmic}
\end{algorithm}

%% file: paper_files/results.tex
\section{Experimental Results}\label{sec:results}

We now present the results of our experimental evaluation of \method{}. We first define the experimental setup, including the target models, tasks, and configurations (\cref{sec:results:setup}), as well as the baseline attacks we compare against (\cref{sec:results:baselines}). We report the main results on both encoder and decoder settings in \cref{sec:results:main}. We study the factors that impact \method{} in \cref{sec:results:ablations}, including initialization (\cref{sec:results:ablations:init}) and quantization (\cref{sec:results:ablations:quant}), with architecture-specific ablations in \cref{sec:results:decoder-ablations} and \cref{sec:results:encoder-ablations} for decoders and encoders, respectively.

\subsection{Experimental Setup}\label{sec:results:setup}

We first summarize the evaluation protocol used throughout our experiments, unless otherwise specified.

\subsubsection{Models and tasks}
We evaluate \method{} on two modern transformer models: \gemma{}~\cite{gemma3} and \embgemma{}~\cite{embeddinggemma}. In the decoder setting, we attack \gemma{} using gradients from a next-token prediction objective, matching the standard causal language-modeling setup used in pretraining and supervised fine-tuning. In the encoder setting, we attack \embgemma{} using gradients from a sequence-classification objective with randomly assigned binary labels. 
Unless otherwise stated, all gradients are computed in FP32, and all attack computations are also performed in FP32.

\subsubsection{Relevant hyperparameters}
For all experiments, unless otherwise specified, we use $\MaxLayers=15$  layers in the span loss, a choice we validate empirically in \cref{sec:results:ablations:layers}. The dummy input contains $\NumInitializations=500$ initializations that are optimized over $\NumSteps=3000$ Adam steps at $\LearningRate=3\times 10^{-2}$ each. We use a \texttt{ReduceLROnPlateau} learning rate scheduler with LR decay of 0.1, LR decay patience of 100, and a min LR of $10^{-6}$. For the encoder track, we set $\BackwardCoeff=0.3$, and for the decoder experiments, we vary $\DedupWeightCoeff$ dependent on the batch size, with $\DedupWeightCoeff=0.1$ for $B=2$, $\DedupWeightCoeff=0.05$ for $B=4$, and $\DedupWeightCoeff=0.0125$ for $B=8$. These values are selected based on preliminary experiments on a small set of validation batches, and we use the same values across all subsequent experiments.

\subsubsection{Datasets}

We use \wikitext{}-103~\cite{wikitext} as the source of client batches, as well as for deriving the initialization prior distribution. We build a pool of 1{,}000 fixed-length passages by retaining individual paragraphs of the training split (excluding headings and blank lines) that tokenize to at least $256$ tokens under the \gemma{} tokenizer (excluding the prepended BOS token), and truncating each to exactly $256$ tokens. To ensure the ground-truth tokens are well defined, we discard passages whose decoded text does not re-tokenize to the identical token sequence. At attack time, each passage is tokenized with the target model tokenizer and truncated to the desired sequence length $\MaxSeqLength \leq 256$. We run each experiment on a set of 10 batches of $B$ sequences, where each batch is sampled from this pool.

\subsubsection{Metrics}

We report token-level ROUGE-1 and ROUGE-L F1 scores~\cite{rouge} between the reconstructed sequences and the original client batch over the set of token IDs. ROUGE-1 measures unigram overlap, while ROUGE-L measures longest-common-subsequence overlap, capturing both token-level recovery and partial sequence-order recovery. Since the attack recovers the client sequences without a canonical ordering within the batch, for batch sizes $B>1$ we align reconstructions with originals via a one-to-one matching: each reconstructed sequence is assigned to a distinct original sequence, and no two reconstructions may be matched to the same original. Among all such matchings, we select the one that maximizes the sum of ROUGE-1 F1 scores over the batch, breaking ties by the sum of ROUGE-L F1 scores, and report ROUGE scores averaged over the matched pairs.

For decoder experiments, we exclude the final input token of each sequence from evaluation for all attacks. Under our next-token prediction setup, the final input position has no in-sequence next-token target, and no later loss term can depend on it through the causal attention path. As a result, its embedding is not identifiable from the gradient signal used in our evaluation.


\subsubsection{Hardware}
All GPU experiments were run on a single NVIDIA H200 SXM5 GPU with 16 CPU cores and 128GB of RAM. Most \dager{} baseline experiments were instead run on 32 CPU cores with 180GB of RAM. This was sufficient because, in the settings where \dager{} fails, it typically terminates early in the recovery stage rather than becoming compute-bound.

\subsection{Baseline Attacks} \label{sec:results:baselines}

We compare \method{} against two strong text-specific gradient inversion attacks: \lamp{}~\cite{lamp} and \dager{}~\cite{dager}. \lamp{} is an optimization-based attack that alternates between continuous and discrete optimization while incorporating a language-model prior. \dager{} is an algebraic attack that exploits low-rank structure in transformer gradients to enable exact or near-exact recovery under certain architectural and batch-size conditions. These methods are strong prior baselines and provide meaningful comparisons across the settings we study.

\subsubsection{LAMP}

We adapt \lamp{} for the encoder setting. Following its original design, \lamp{} uses a language-model prior during optimization. We use \gemma{} as this prior when attacking \embgemma{}, giving \lamp{} a stronger prior compared to the \gpttwo{} prior used in the original \lamp{} evaluation, and without needing to adapt the tokenization. Because \lamp{} is sensitive to hyperparameters, particularly the perplexity coefficient, learning-rate decay, and gradient-matching loss variant, we tune these parameters before running the main experiments. Specifically, we consider initializations of the form $\bar{Z}_0 = Z_0 + Z_\sigma$, where $Z_\sigma \sim \mathcal{N}(0, \sigma^2)$, representing the ground-truth initialization $Z_0$ with added Gaussian noise. We increase $\sigma$ until \lamp{}'s ROUGE-1 score drops to $0$. At the largest value of $\sigma$ for which \lamp{} still achieves nonzero ROUGE-1, we perform a grid search over the relevant hyperparameters and select the configuration with the highest mean ROUGE-1. We then use this fixed configuration for all subsequent \lamp{} experiments.

For the decoder setting, \lamp{} is not directly applicable because it assumes knowledge of the labels. In next-token prediction, these labels are the shifted input tokens and would directly leak the input sequence. We nevertheless attempted a label-free adaptation of \lamp{}, but found that it failed to achieve nonzero ROUGE-1 even in the $B=1$ setting. We therefore do not include \lamp{} in the reported decoder tables.

\subsubsection{DAGER}

For the decoder setting, we adapt \dager{} to the models used in our evaluation. We follow the original paper by setting the first-layer filtering threshold to $10^{-5}$, but we relax the second-layer threshold to $10^{-1}$. We found that smaller second-layer thresholds were too strict in our setting and removed valid candidates. In the undefended decoder setting, this relaxed threshold still allows \dager{} to recover efficiently, as shown in \cref{sec:results:defended}.

We do not include \dager{} in the encoder setting. The \embgemma{} encoder uses RoPE-style positional embeddings, which differ from the positional-embedding assumptions in the original \dager{} setup. In particular, because RoPE is applied after the query and key projections, a direct adaptation of \dager{} can recover a bag of tokens for the sequence but not their positions. Recovering ordered batches would then require enumerating possible assignments of the recovered $BT$ tokens into $B$ ordered sequences of length $T$. This search space is on the order of $\frac{(BT)!}{B!}$. Even for $B=2$ and $T=16$, this is approximately $\frac{32!}{2!} \approx 1.3 \times 10^{35}$ possible enumerated batches, making direct enumeration computationally infeasible.

\subsection{Main Results} \label{sec:results:main}

We now present the main reconstruction results for \method{}. We first evaluate robustness to additive gradient noise, then study how reconstruction quality changes as the batch size and sequence length vary.

\subsubsection{\method{} on noisy gradients} \label{sec:results:defended}

\begin{figure}[!t]
\centering
\includegraphics[width=0.85\linewidth]{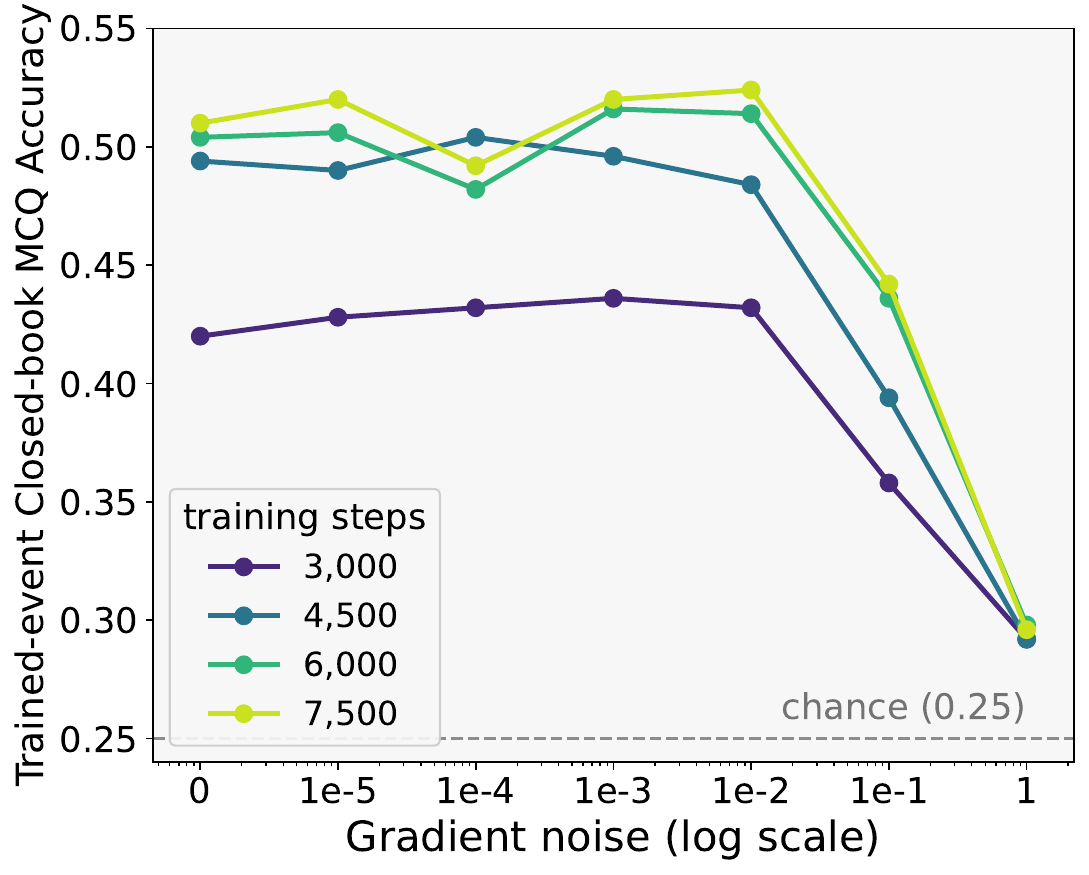}
\caption{Closed-book MCQ accuracy of \gemma{} fine-tuned on FictionalQA.}
\label{fig:dp-utility-decoder-sgd}
\end{figure}

Following \lamp{}~\cite{lamp}, we evaluate reconstruction under additive Gaussian gradient noise. Before the attacker observes the gradients, each gradient element is independently perturbed by noise drawn from $\mathcal{N}(0, \sigma^2)$.  For this experiment, we fix the sequence length to $\MaxSeqLength=16$ for all sequences in the batch.

To select noise levels that preserve training utility, we first fine-tune \gemma{} on FictionalQA~\cite{fictionalqa} under different gradient-noise scales and measure closed-book $4$-way MCQ accuracy, where an answer is correct if the most likely token after the prompt matches the ground-truth answer. As shown in \cref{fig:dp-utility-decoder-sgd}, accuracy is largely unaffected until $\sigma$ exceeds $10^{-2}$ and degrades towards random chance by $\sigma=1$. We therefore evaluate attacks in the utility-preserving range $\sigma \in [0, 10^{-3}]$.

\input{paper_files/generated/main-sigma-b-decoder}

We report decoder results in \cref{tab:main-sigma-b-decoder}. In this setting, \method{} is substantially more robust to additive noise than \dager{}. In the undefended case, both attacks recover high-quality reconstructions, although \method{} achieves slightly higher ROUGE across the evaluated batch sizes. Once any Gaussian noise is added, however, \dager{} collapses to zero ROUGE, whereas \method{} continues to recover meaningful token overlap. Reconstruction quality decreases as both $\sigma$ and $B$ increase, but \method{} remains the only evaluated decoder attack with nonzero recovery under noisy gradients.

\input{paper_files/generated/main-sigma-b-encoder}

On the other hand, in the encoder setting, as shown in \cref{tab:main-sigma-b-encoder}, \method{} substantially outperforms \lamp{} in the undefended case across all evaluated batch sizes. However, encoder recovery is more sensitive to noise than decoder recovery. At $\sigma=10^{-5}$, \method{}'s ROUGE scores drop sharply, especially for larger batches, and the gap to \lamp{} narrows. This suggests that the joint input optimization provides a weaker signal than token-by-token recovery, as the former is more vulnerable to noisy subspace estimates.

\subsubsection{\method{} across batch size--sequence length tradeoffs} \label{sec:results:batch-seq-tradeoffs}

We next evaluate how reconstruction quality changes as the total number of client tokens increases. We vary the batch size $B$ and per-sequence length $\MaxSeqLength$, and report results as a function of the total number of non-padding tokens $\NumTokens = B\MaxSeqLength$. For decoder experiments, we use $\sigma=5 \times 10^{-5}$, while for encoder experiments, we use $\sigma=0$ to isolate scaling behavior without additive noise.

\input{paper_files/generated/main-batch-seq-decoder}

For decoder models, \method{} achieves nonzero and often high ROUGE across all evaluated batch size--sequence length configurations, while \dager{} fails due to the added noise. Moreover, as expected, reconstruction quality generally decreases as $\NumTokens$ increases. For a fixed total token budget, recovery tends to be easier with fewer, longer sequences than with larger batches of shorter sequences, consistent with the sequential decoder attack benefiting from an already recovered prefix.

\input{paper_files/generated/main-batch-seq-encoder}

For encoder models, \method{} consistently outperforms \lamp{} across the evaluated configurations. ROUGE-1 remains comparatively strong even at larger token counts, whereas ROUGE-L decreases more sharply as sequences become longer. This indicates that \method{} often recovers many correct tokens, but that these tokens are less consistently placed in the correct order. This behavior reflects the difficulty of joint ordered recovery in non-causal encoders.

Overall, these results show that \method{} is the strongest evaluated attack on noisy decoder gradients and scales substantially better than prior optimization-based attacks in undefended encoder settings.
\subsection{Ablation Studies} \label{sec:results:ablations}

We now ablate the main design choices of \method{}. Unless otherwise specified, we use $B=4$. For decoder ablations, we set $\MaxSeqLength=16$ and $\sigma=5\times 10^{-5}$. For encoder ablations, we set $\MaxSeqLength=32$ and $\sigma=0$.

\subsubsection{Impact of initialization strategy} \label{sec:results:ablations:init}
Initialization provides an important prior for the non-convex embedding optimization. We compare the default position-dependent WikiText Gaussian prior against alternative initialization strategies in \cref{tab:init-table}.

\input{paper_files/generated/init-table}

Fitting the same position-dependent Gaussian prior on IMDb~\cite{imdb} instead of WikiText yields nearly identical performance in both settings, suggesting that the specific public corpus used to estimate the prior is not the main factor. In contrast, sampling from a position-agnostic distribution over the vocabulary embedding space reduces reconstruction quality, indicating that position-specific statistics provide useful information. Finally, independently sampling embedding coordinates degrades decoder performance substantially and also underperforms the structured prior for encoders. These results highlight the importance of preserving realistic embedding-space structure during initialization.

\subsubsection{\method{} under lower precision} \label{sec:results:ablations:quant}

All experiments so far have used FP32 gradients. We next test whether \method{} remains effective when the observed gradients are stored at lower precision. To isolate numerical precision from additive noise, we set $\sigma=0$ for the decoder experiments in this ablation. With BF16 gradients, \method{}'s ROUGE-1 decreases only modestly, from $99.1\%$ to $94.5$\% for the decoder and from $72.2\%$ to $65.5\%$ for the encoder. In contrast, \dager{} drops to zero ROUGE-1, because the reduced precision makes its threshold-based span-membership tests unreliable. This supports the advantage of \method{}'s continuous multi-layer objective, which can aggregate weak subspace signals across layers rather than relying on brittle discrete filtering.

\subsubsection{\method{} using different numbers of layers} \label{sec:results:ablations:layers}
\begin{figure*}[!t]
\centering
\begin{minipage}[t]{0.49\textwidth}
\centering
\includegraphics[width=\textwidth]{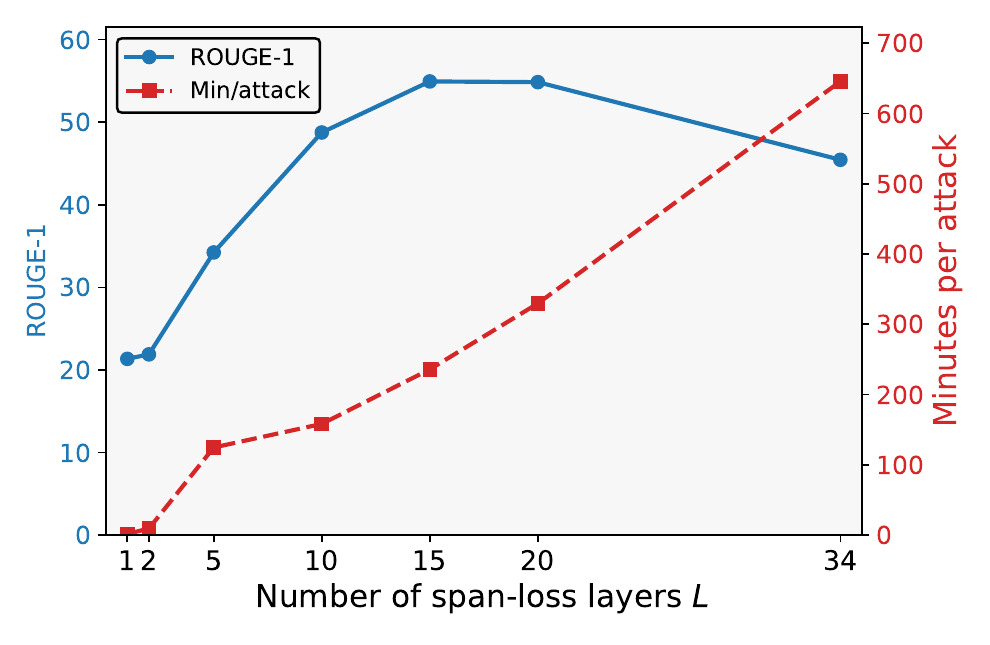}
\captionof{figure}{ROUGE-1 and mean per-attack wall-clock as a function of the number of span-loss layers $L$ for \gemma{}.}
\label{fig:layer-ablations-decoder}
\end{minipage}
\hfill
\begin{minipage}[t]{0.49\textwidth}
\centering
\includegraphics[width=\textwidth]{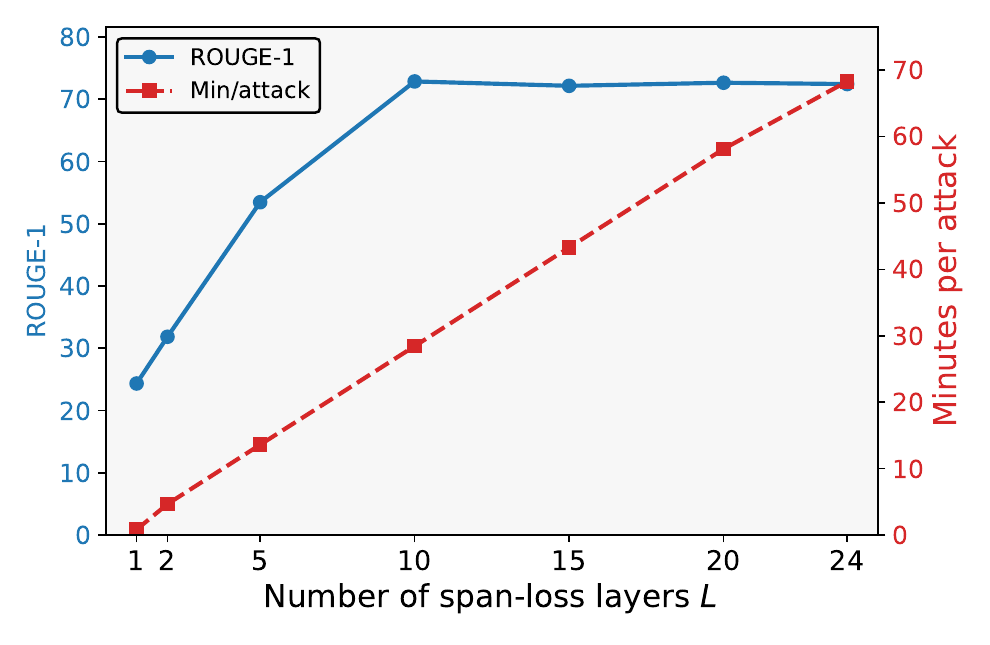}
\captionof{figure}{ROUGE-1 and mean per-attack wall-clock as a function of the number of span-loss layers $L$ for \embgemma{}.}
\label{fig:layer-ablations-encoder}
\end{minipage}
\end{figure*}

The number of span-loss layers $\MaxLayers$ controls a tradeoff between reconstruction signal and computational cost. Using more layers provides additional subspace constraints and can stabilize optimization, but it also increases the cost of each forward and backward pass. We vary $\MaxLayers$ and report both ROUGE-1 and mean per-attack wall-clock time in \cref{fig:layer-ablations-decoder,fig:layer-ablations-encoder} for the decoder and encoder settings, respectively. In both settings, reconstruction improves rapidly for the first few layers and then saturates around $10$--$15$ layers, while runtime grows approximately linearly. We therefore use $\MaxLayers=15$ as the default configuration.

\subsubsection{Recovering encoder sequences without backward loss} \label{sec:results:encoder-ablations}

We next ablate the backward span-distance loss in the encoder setting. Removing this term reduces ROUGE-1 from $72.2\%$ to $41.4\%$ and ROUGE-L from $54.1\%$ to $36.8\%$. Qualitatively, the failures are caused primarily by duplicate reconstructions across batch elements rather than by repetition within individual sequences. Without the backward loss, none of the $10$ attacks recovers four distinct sequences, and in $3$ of the $10$ attacks all four reconstructions collapse to the same batch member. This supports the role of the backward loss in preventing cross-example collapse by encouraging the optimized hidden states to cover the full gradient-induced subspace.

\subsubsection{Decoder-specific ablations} \label{sec:results:decoder-ablations}

Finally, we ablate the decoder-specific components of \method{} in \cref{tab:decoder-abl}.

\input{paper_files/tables/decoder-abl}

First, we replace the sequential causal recovery procedure with joint optimization of the full input embedding matrix, using the backward loss as in the encoder attack described in \cref{sec:methodology:encoder}. This reduces both ROUGE-1 and ROUGE-L by nearly $25$ points, indicating that explicitly exploiting the causal structure leads to substantially more stable recovery in the decoder setting.

Second, we remove the first-token deduplication loss described in \cref{sec:methodology:decoder}. This reduces ROUGE by roughly $20$ points, as without deduplication, multiple starting tokens often converge to the same solution. Because all subsequent tokens in a decoder sequence depend causally on the first token, this early duplication prevents accurate recovery of later tokens in the affected sequences.

%% file: paper_files/generated/main-sigma-b-decoder.tex
\begin{table*}[t]
\centering
\caption{Main decoder results for \method{} on \gemma{} across different noise levels $\sigma$ and batch sizes $B$.}
\label{tab:main-sigma-b-decoder}
\small
\setlength{\tabcolsep}{3.5pt}
\resizebox{\textwidth}{!}{%
\begin{tabular}{@{}cl cccccccccc@{}}
\toprule
\multirow{2}{*}{$B$} & \multirow{2}{*}{Attack}
& \multicolumn{2}{c}{$\sigma{} = 0$}
& \multicolumn{2}{c}{$\sigma{} = 10^{-5}$}
& \multicolumn{2}{c}{$\sigma{} = 5 \times 10^{-5}$}
& \multicolumn{2}{c}{$\sigma{} = 10^{-4}$}
& \multicolumn{2}{c}{$\sigma{} = 10^{-3}$} \\
\cmidrule(lr){3-4}\cmidrule(lr){5-6}\cmidrule(lr){7-8}\cmidrule(lr){9-10}\cmidrule(lr){11-12}
& & ROUGE-1 & ROUGE-L & ROUGE-1 & ROUGE-L & ROUGE-1 & ROUGE-L & ROUGE-1 & ROUGE-L & ROUGE-1 & ROUGE-L \\
\midrule
\multirow{2}{*}{$1$} & \dager{} & 93.7 & 93.7 & 0.0 & 0.0 & 0.0 & 0.0 & 0.0 & 0.0 & 0.0 & 0.0 \\
& \method{} & \textbf{98.8} & \textbf{98.8} & \textbf{97.4} & \textbf{97.4} & \textbf{95.9} & \textbf{95.9} & \textbf{94.1} & \textbf{94.1} & \textbf{71.1} & \textbf{67.8} \\
\midrule
\multirow{2}{*}{$2$} & \dager{} & 92.2 & 92.2 & 0.0 & 0.0 & 0.0 & 0.0 & 0.0 & 0.0 & 0.0 & 0.0 \\
& \method{} & \textbf{98.9} & \textbf{98.9} & \textbf{89.8} & \textbf{89.4} & \textbf{63.6} & \textbf{63.1} & \textbf{49.6} & \textbf{49.6} & \textbf{38.1} & \textbf{35.8} \\
\midrule
\multirow{2}{*}{$4$} & \dager{} & 90.6 & 90.4 & 0.0 & 0.0 & 0.0 & 0.0 & 0.0 & 0.0 & 0.0 & 0.0 \\
& \method{} & \textbf{99.1} & \textbf{99.1} & \textbf{92.9} & \textbf{92.9} & \textbf{54.9} & \textbf{54.3} & \textbf{42.2} & \textbf{41.6} & \textbf{27.8} & \textbf{26.0} \\
\midrule
\multirow{2}{*}{$8$} & \dager{} & 85.8 & 85.7 & 0.0 & 0.0 & 0.0 & 0.0 & 0.0 & 0.0 & 0.0 & 0.0 \\
& \method{} & \textbf{92.2} & \textbf{92.2} & \textbf{83.9} & \textbf{83.9} & \textbf{39.7} & \textbf{38.9} & \textbf{27.8} & \textbf{26.7} & \textbf{19.4} & \textbf{17.8} \\
\bottomrule
\end{tabular}%
}
\end{table*}

%% file: paper_files/generated/main-sigma-b-encoder.tex
\begin{table}[t]
\centering
\caption{Main encoder results for \method{} on \embgemma{} across different noise levels $\sigma$ and batch sizes $B$.}
\label{tab:main-sigma-b-encoder}
\small
\setlength{\tabcolsep}{5pt}
\resizebox{\columnwidth}{!}{%
\begin{tabular}{@{}cl cccc@{}}
\toprule
\multirow{2}{*}{$B$} & \multirow{2}{*}{Attack}
& \multicolumn{2}{c}{$\sigma{} = 0$}
& \multicolumn{2}{c}{$\sigma{} = 10^{-5}$} \\
\cmidrule(lr){3-4}\cmidrule(lr){5-6}
& & ROUGE-1 & ROUGE-L & ROUGE-1 & ROUGE-L \\
\midrule
\multirow{2}{*}{$1$} & \lamp{} & 9.8 & 9.8 & 5.9 & 5.9 \\
& \method{} & \textbf{90.8} & \textbf{86.7} & \textbf{48.1} & \textbf{41.4} \\
\midrule
\multirow{2}{*}{$2$} & \lamp{} & 15.2 & 12.6 & 17.1 & 14.9 \\
& \method{} & \textbf{87.8} & \textbf{75.9} & \textbf{27.9} & \textbf{24.1} \\
\midrule
\multirow{2}{*}{$4$} & \lamp{} & 16.0 & 12.2 & 16.2 & 12.8 \\
& \method{} & \textbf{80.9} & \textbf{66.6} & \textbf{20.4} & \textbf{17.7} \\
\midrule
\multirow{2}{*}{$8$} & \lamp{} & 18.5 & 15.2 & 19.1 & 15.6 \\
& \method{} & \textbf{71.0} & \textbf{56.7} & \textbf{21.3} & \textbf{16.8} \\
\midrule
\multirow{2}{*}{$16$} & \lamp{} & 23.3 & 17.4 & \textbf{22.9} & \textbf{18.3} \\
& \method{} & \textbf{60.3} & \textbf{50.7} & 22.3 & 17.5 \\
\bottomrule
\end{tabular}%
}
\end{table}

%% file: paper_files/generated/main-batch-seq-decoder.tex
\begin{table}[t]
\centering
\caption{Decoder reconstruction results for \gemma{} across the total token count $T$ and batch size $B$ at $\sigma{=}\num{5e-5}$.}
\label{tab:decoder-t-b}
\small
\setlength{\tabcolsep}{4pt}
\resizebox{\columnwidth}{!}{%
\begin{tabular}{@{}cc*{4}{c}@{}}
\toprule
\multirow{2}{*}{$T$}
& \multirow{2}{*}{$B$}
& \multicolumn{2}{c}{\dager{}}
& \multicolumn{2}{c}{\method{}} \\
\cmidrule(lr){3-4}
\cmidrule(lr){5-6}
& & ROUGE-1 & ROUGE-L & ROUGE-1 & ROUGE-L \\
\midrule

\multirow{1}{*}{$16$}
& $1$ & 0.0 & 0.0 & \textbf{95.9} & \textbf{95.9} \\
\midrule

\multirow{3}{*}{$32$}
& $1$ & 0.0 & 0.0 & \textbf{98.6} & \textbf{98.6} \\
& $2$ & 0.0 & 0.0 & \textbf{63.6} & \textbf{63.1} \\
& $4$ & 0.0 & 0.0 & \textbf{58.8} & \textbf{58.4} \\
\midrule

\multirow{4}{*}{$64$}
& $1$ & 0.0 & 0.0 & \textbf{94.6} & \textbf{94.4} \\
& $2$ & 0.0 & 0.0 & \textbf{62.7} & \textbf{62.5} \\
& $4$ & 0.0 & 0.0 & \textbf{54.9} & \textbf{54.3} \\
& $8$ & 0.0 & 0.0 & \textbf{57.9} & \textbf{57.7} \\
\midrule

\multirow{4}{*}{$128$}
& $1$ & 0.0 & 0.0 & \textbf{91.2} & \textbf{89.4} \\
& $2$ & 0.0 & 0.0 & \textbf{58.6} & \textbf{56.7} \\
& $4$ & 0.0 & 0.0 & \textbf{42.3} & \textbf{40.7} \\
& $8$ & 0.0 & 0.0 & \textbf{39.7} & \textbf{38.9} \\
\bottomrule
\end{tabular}%
}
\end{table}

%% file: paper_files/generated/main-batch-seq-encoder.tex
\begin{table}[t]
\centering
\caption{Encoder reconstruction results for \embgemma{} across the total token count $T$ and batch size $B$ at $\sigma{=}0$.}
\label{tab:encoder-t-b}
\small
\setlength{\tabcolsep}{5pt}
\resizebox{\columnwidth}{!}{%
\begin{tabular}{@{}cc*{4}{c}@{}}
\toprule
\multirow{2}{*}{$T$}
& \multirow{2}{*}{$B$}
& \multicolumn{2}{c}{\lamp{}}
& \multicolumn{2}{c}{\method{}} \\
\cmidrule(lr){3-4}
\cmidrule(lr){5-6}
& & ROUGE-1 & ROUGE-L & ROUGE-1 & ROUGE-L \\
\midrule

\multirow{1}{*}{$16$}
& $1$ & 9.8 & 9.8 & \textbf{90.8} & \textbf{86.7} \\
\midrule

\multirow{3}{*}{$32$}
& $1$ & 9.6 & 7.5 & \textbf{84.6} & \textbf{75.1} \\
& $2$ & 15.2 & 12.6 & \textbf{87.8} & \textbf{75.9} \\
& $4$ & 12.7 & 10.8 & \textbf{83.9} & \textbf{79.7} \\
\midrule

\multirow{4}{*}{$64$}
& $1$ & 17.6 & 12.7 & \textbf{83.3} & \textbf{56.2} \\
& $2$ & 13.8 & 12.0 & \textbf{79.0} & \textbf{59.5} \\
& $4$ & 16.0 & 12.2 & \textbf{80.9} & \textbf{66.6} \\
& $8$ & 20.3 & 18.6 & \textbf{80.9} & \textbf{74.2} \\
\midrule

\multirow{4}{*}{$128$}
& $1$ & 22.9 & 15.7 & \textbf{82.1} & \textbf{43.8} \\
& $2$ & 18.0 & 12.6 & \textbf{79.4} & \textbf{52.2} \\
& $4$ & 19.1 & 13.9 & \textbf{72.2} & \textbf{54.1} \\
& $8$ & 18.5 & 15.2 & \textbf{71.0} & \textbf{56.7} \\
\midrule

\multirow{5}{*}{$256$}
& $1$ & 26.8 & 17.0 & \textbf{64.5} & \textbf{26.5} \\
& $2$ & 22.1 & 15.8 & \textbf{68.7} & \textbf{35.8} \\
& $4$ & 23.1 & 15.7 & \textbf{66.2} & \textbf{38.0} \\
& $8$ & 21.8 & 16.9 & \textbf{57.9} & \textbf{41.8} \\
& $16$ & 23.3 & 17.4 & \textbf{60.3} & \textbf{50.7} \\
\bottomrule
\end{tabular}%
}
\end{table}

%% file: paper_files/generated/init-table.tex
\begin{table}[t]
\centering
\caption{Impact of different initialization strategies on \method{}. We report ROUGE-1 and ROUGE-L, denoted as R-1 and R-L, respectively.}
\label{tab:init-table}
\small
\setlength{\tabcolsep}{8pt}
\begin{tabular}{@{}llcc@{}}
\toprule
Model & Configuration & R-1 & R-L \\
\midrule

\multirow{4}{*}{\gemma{}}
& Default & 54.9 & \textbf{54.3} \\
& IMDb Init. & \textbf{55.4} & 54.0 \\
& Vocabulary Init. & 46.0 & 44.9 \\
& Random Init. & 24.3 & 24.3 \\

\midrule

\multirow{4}{*}{\embgemma{}}
& Default & \textbf{72.2} & \textbf{54.1} \\
& IMDb Init. & 71.9 & 53.5 \\
& Vocabulary Init. & 67.4 & 47.2 \\
& Random Init. & 68.8 & 49.9 \\

\bottomrule
\end{tabular}
\end{table}

%% file: paper_files/tables/decoder-abl.tex

\begin{table}[t]
\centering
\caption{Decoder attack configuration ablation.}
\label{tab:decoder-abl}
\small
\setlength{\tabcolsep}{8pt}
\begin{tabular}{@{}lcc@{}}
\toprule
Configuration & ROUGE-1 & ROUGE-L \\
\midrule

 Default & \textbf{54.9} & \textbf{54.3} \\
 W/o First-Token Deduplication & 32.0 & 31.7 \\
 W/ Backward Loss & 30.8 & 29.5 \\
\bottomrule
\end{tabular}
\end{table}

%% file: paper_files/limitations.tex
\section{Limitations and Future Work} \label{sec:limitations}

While \method{} demonstrates an effective and robust approach to gradient inversion for transformer LLMs, several limitations suggest natural directions for future work.

First, \method{} builds on the low-rank structure of linear-layer gradients~\cite{dager,grain,spear,spear++}. This structure is most informative when the number of token representations contributing to a gradient remains smaller than the relevant hidden dimension. As the total number of tokens $T$ approaches or exceeds the hidden dimension $d$, the corresponding gradient subspace becomes full-rank, reducing or eliminating the discriminative power of the subspace objective.

Second, although our results show that \method{} remains effective under additive DP-style noise, our evaluation does not exhaust the space of possible privacy-preserving defenses. In particular, different clipping strategies~\cite{clipping,adaptiveclipping,sampleadaptiveclipping}, gradient masking~\cite{communicationstrategies}, secure aggregation~\cite{secureaggregation}, compression~\cite{compression}, or combinations of these mechanisms may affect the available subspace signal in different ways. A more complete characterization of how such defenses interact with \method{} would be valuable, both for understanding the limits of the reconstruction attack and for designing stronger defenses.

Finally, \method{} focuses on recovering token-level information through continuous embedding optimization and therefore does not always produce an exact ordered reconstruction of the original batch. In many cases, however, most recovered sequences contain substantial portions of the original content, leaving room for hybrid attacks that combine \method{}'s continuous subspace recovery with discrete refinement procedures. For example, a language-model prior could be used to reorder recovered tokens or infer missing ones, similar to the refinement used in LAMP~\cite{lamp}. We view such reconstruction pipelines as a promising direction for future work.

%% file: paper_files/conclusion.tex
\section{Conclusion} \label{sec:conclusion}

We introduced \method{}, a continuous subspace-based gradient inversion attack for transformer language models. \method{} turns the low-rank structure of transformer linear-layer gradients into a differentiable embedding-recovery objective by directly optimizing token embeddings to align their hidden representations with gradient-induced subspaces. This formulation supports both decoder-only and encoder-only models, using causal sequential recovery for decoders and a bidirectional alignment objective for encoders. Across our experiments, this continuous subspace view substantially expands the practical threat posed by gradient inversion. On decoder-only models, \method{} remains competitive in undefended settings and continues to recover meaningful text where prior attacks fail under additive gradient noise or reduced numerical precision. On encoder-only models, where causal token-by-token recovery is unavailable, \method{} substantially outperforms prior optimization-based attacks and scales to larger batched updates. These findings show that transformer LLM gradients can retain significant textual information even after common forms of perturbation or quantization. In particular, modest gradient noise or lower-precision gradient computation should not be assumed to eliminate reconstruction risk. This reinforces the need for carefully evaluated and applied privacy defenses in practical federated learning deployments.

%% file: paper_files/acknowledgements.tex
\section*{Acknowledgments}

Parts of this work were done at INSAIT, Sofia University “St. Kliment Ohridski”, Bulgaria. This work was partially funded by the Ministry of Education and Science of Bulgaria (support for INSAIT, part of the Bulgarian National Roadmap for Research Infrastructure). This project was supported with computational resources provided by Google Cloud Platform (GCP).

%% file: paper_files/appendix.tex
\appendices
\section{Basis Rotation for First-Token Decoder Deduplication}\label{sec:dedup-projs}

In decoder-only models, the first token of each sequence is especially important, because every later token representation in that sequence causally depends on it. When attacking batches with $\BatchSize>1$, however, the span constraints for the first token can admit repeated solutions, as all correct first-token embeddings are equally valid global minima of the forward loss. To reduce this failure mode, \textsc{DedupProjs} constructs, for each layer $l$, a projection matrix onto the directions of the gradient-induced subspace $\GradientSubspace{l}$ that are already explained by the $b-1$ previously recovered first tokens.

The algorithm first stacks the previously recovered first-token embeddings,
\[
\RecoveredInput{0}_{\mathrm{prev}} \leftarrow
\left[
\RecoveredInput{0}_{1,1};
\RecoveredInput{0}_{2,1};
\cdots;
\RecoveredInput{0}_{b-1,1}
\right].
\]
It then forwards these embeddings through the attacked transformer layers, obtaining $\RecoveredInput{l}_{\mathrm{prev}}$ at each layer. Given an orthonormal basis $\SubspaceBasis{l}\in\mathbb{R}^{\NumTokens\times d}$ of the gradient-induced subspace $\GradientSubspace{l}$, the matrix
\[
\bm{C}^{l}_{:,1:b-1} \leftarrow
\SubspaceBasis{l}
\left(
\RecoveredInput{l}_{\mathrm{prev}}
\right)^\top
\]
contains the coordinates of the previously recovered hidden states in the basis $\SubspaceBasis{l}$. Thus, $\bm{C}^{l}$ records how the projections of the previous first-token hidden states decompose along the basis directions of $\GradientSubspace{l}$. Each of the first $b-1$ columns corresponds to one previously recovered first token, while each row corresponds to one basis direction of $\GradientSubspace{l}$.

The SVD of $\bm{C}^{l}$ gives an orthogonal rotation $\bm{R}^{l}$ of the coordinates inside the gradient-induced subspace. We use this rotation to form a rotated basis
\[
\RotatedSubspaceBasis{l}
\leftarrow
\left(
\bm{R}^{l}
\right)^\top
\SubspaceBasis{l}.
\]
This full rotated basis still spans the same subspace $\GradientSubspace{l}$, but its leading coordinates are aligned with the directions that explain the previously recovered first-token hidden states. For deduplication, we therefore do not use the full rotated basis. Instead, when recovering the first token of sequence $b$, we keep only the first $b-1$ rotated basis vectors:
\[
\RotatedSubspaceBasis{l,b}
\leftarrow
\left(
\RotatedSubspaceBasis{l}
\right)_{1:b-1,:}.
\]
The previous-token projector is then defined as
\[
\SubspaceDedupProjection{l,b}
\leftarrow
\left(
\RotatedSubspaceBasis{l,b}
\right)^\top
\RotatedSubspaceBasis{l,b}.
\]
Thus, $\SubspaceDedupProjection{l,b}$ projects only onto the portion of the gradient-induced subspace $\GradientSubspace{l}$ associated with the previously recovered first tokens, rather than onto the entire gradient-induced subspace $\GradientSubspace{l}$.

The resulting projector $\SubspaceDedupProjection{l,b}$ is used in the decoder objective to penalize overlap between the new first-token hidden state and the directions already used by earlier first tokens:
\[
\DedupLoss
\leftarrow
\sum_{l=1}^{\MaxLayers}
\left\|
\frac{
	\OptimizedInput{l}_{b,1}
}{
	\left\|
	\OptimizedInput{l}_{b,1}
	\right\|_2
}
(\SubspaceDedupProjection{l,b})^\top
\right\|^2_2 .
\]
This encourages different batch elements to use distinct first-token directions, while the standard forward span loss still keeps the recovered representation inside the full gradient-induced subspace $\GradientSubspace{l}$.

\begin{algorithm}[t]
	\caption{\textsc{DedupProjs}: Basis Rotation for Decoder First-Token Deduplication}
	\label{alg:dedup-projs}
	\begin{algorithmic}[1]
		\REQUIRE Transformer blocks $\{\TransformerBlock{l}\}_{l=0}^{\MaxLayers-1}$,\\
		\, Orthonormal subspace bases $\{\SubspaceBasis{l}\}_{l=1}^{\MaxLayers}$,\\
		\, Previously recovered first-token embeddings $\RecoveredInput{0}_{1:b-1,1}$,\\
		\, Attention mask $\AttentionMask$
		\ENSURE Previous-token projections $\{\SubspaceDedupProjection{l,b}\}_{l=1}^{\MaxLayers}$
		
		\STATE $\RecoveredInput{0}_{\mathrm{prev}} \leftarrow
		\left[
		\RecoveredInput{0}_{1,1};
		\RecoveredInput{0}_{2,1};
		\cdots;
		\RecoveredInput{0}_{b-1,1}
		\right]$
		
		\FOR{$l=1$ to $\MaxLayers$}
		\STATE $\bm{C}^{l} \leftarrow
		\bm{0}_{\NumTokens\times \max(\NumTokens,b-1)}$
		\STATE $\RecoveredInput{l}_{\mathrm{prev}} \leftarrow
		\TransformerBlock{l-1}
		\left(
		\RecoveredInput{l-1}_{\mathrm{prev}};
		\AttentionMask
		\right)$
		
		\STATE $\bm{C}^{l}_{:,1:b-1} \leftarrow
		\SubspaceBasis{l}
		\left(
		\RecoveredInput{l}_{\mathrm{prev}}
		\right)^\top$
		
		\STATE $\bm{R}^{l}, \_, \_
		\leftarrow
		\operatorname{SVD}
		\left(
		\bm{C}^{l}
		\right)$
		
		\STATE $\RotatedSubspaceBasis{l}
		\leftarrow
		\left(
		\bm{R}^{l}
		\right)^\top
		\SubspaceBasis{l}$
		
		\STATE $\RotatedSubspaceBasis{l,b}
		\leftarrow
		(\RotatedSubspaceBasis{l})_{1:b-1,:}$
		
		\STATE $\SubspaceDedupProjection{l,b}
		\leftarrow
		\left(
		\RotatedSubspaceBasis{l,b}
		\right)^\top
		\RotatedSubspaceBasis{l,b}$
		\ENDFOR
		
		\RETURN $\{\SubspaceDedupProjection{l,b}\}_{l=1}^{\MaxLayers}$
	\end{algorithmic}
\end{algorithm}

%% file: references.bib
@inproceedings{dlg,
  author       = {Ligeng Zhu and
                  Zhijian Liu and
                  Song Han},
  editor       = {Hanna M. Wallach and
                  Hugo Larochelle and
                  Alina Beygelzimer and
                  Florence d'Alch{\'{e}}{-}Buc and
                  Emily B. Fox and
                  Roman Garnett},
  title        = {Deep Leakage from Gradients},
  booktitle    = {Advances in Neural Information Processing Systems 32: Annual Conference
                  on Neural Information Processing Systems 2019, NeurIPS 2019, December
                  8-14, 2019, Vancouver, BC, Canada},
  pages        = {14747--14756},
  year         = {2019},
  url          = {https://proceedings.neurips.cc/paper/2019/hash/60a6c4002cc7b29142def8871531281a-Abstract.html},
  bibsource    = {dblp computer science bibliography, https://dblp.org}
}

@inproceedings{geiping,
  author       = {Jonas Geiping and
                  Hartmut Bauermeister and
                  Hannah Dr{\"{o}}ge and
                  Michael Moeller},
  editor       = {Hugo Larochelle and
                  Marc'Aurelio Ranzato and
                  Raia Hadsell and
                  Maria{-}Florina Balcan and
                  Hsuan{-}Tien Lin},
  title        = {Inverting Gradients - How easy is it to break privacy in federated
                  learning?},
  booktitle    = {Advances in Neural Information Processing Systems 33: Annual Conference
                  on Neural Information Processing Systems 2020, NeurIPS 2020, December
                  6-12, 2020, virtual},
  year         = {2020},
  url          = {https://proceedings.neurips.cc/paper/2020/hash/c4ede56bbd98819ae6112b20ac6bf145-Abstract.html},
  bibsource    = {dblp computer science bibliography, https://dblp.org}
}

@inproceedings{tag,
  author       = {Jieren Deng and
                  Yijue Wang and
                  Ji Li and
                  Chenghong Wang and
                  Chao Shang and
                  Hang Liu and
                  Sanguthevar Rajasekaran and
                  Caiwen Ding},
  editor       = {Marie{-}Francine Moens and
                  Xuanjing Huang and
                  Lucia Specia and
                  Scott Wen{-}tau Yih},
  title        = {{TAG:} Gradient Attack on Transformer-based Language Models},
  booktitle    = {Findings of the Association for Computational Linguistics: {EMNLP}
                  2021, Virtual Event / Punta Cana, Dominican Republic, 16-20 November,
                  2021},
  series       = {Findings of {ACL}},
  pages        = {3600--3610},
  publisher    = {Association for Computational Linguistics},
  year         = {2021},
  url          = {https://doi.org/10.18653/v1/2021.findings-emnlp.305},
  doi          = {10.18653/V1/2021.FINDINGS-EMNLP.305},
  bibsource    = {dblp computer science bibliography, https://dblp.org}
}

@inproceedings{lamp,
  author       = {Mislav Balunovic and
                  Dimitar I. Dimitrov and
                  Nikola Jovanovic and
                  Martin T. Vechev},
  editor       = {Sanmi Koyejo and
                  S. Mohamed and
                  A. Agarwal and
                  Danielle Belgrave and
                  K. Cho and
                  A. Oh},
  title        = {{LAMP:} Extracting Text from Gradients with Language Model Priors},
  booktitle    = {Advances in Neural Information Processing Systems 35: Annual Conference
                  on Neural Information Processing Systems 2022, NeurIPS 2022, New Orleans,
                  LA, USA, November 28 - December 9, 2022},
  year         = {2022},
  url          = {http://papers.nips.cc/paper\_files/paper/2022/hash/32375260090404f907ceae19f3564a7e-Abstract-Conference.html},
  bibsource    = {dblp computer science bibliography, https://dblp.org}
}

@inproceedings{dager,
  author       = {Ivo Petrov and
                  Dimitar I. Dimitrov and
                  Maximilian Baader and
                  Mark Niklas M{\"{u}}ller and
                  Martin T. Vechev},
  editor       = {Amir Globersons and
                  Lester Mackey and
                  Danielle Belgrave and
                  Angela Fan and
                  Ulrich Paquet and
                  Jakub M. Tomczak and
                  Cheng Zhang},
  title        = {{DAGER:} Exact Gradient Inversion for Large Language Models},
  booktitle    = {Advances in Neural Information Processing Systems 37: Annual Conference
                  on Neural Information Processing Systems 2024, NeurIPS 2024, Vancouver,
                  BC, Canada, December 10 - 15, 2024},
  year         = {2024},
  url          = {http://papers.nips.cc/paper\_files/paper/2024/hash/9ff1577a1f8308df1ccea6b4f64a103f-Abstract-Conference.html},
  bibsource    = {dblp computer science bibliography, https://dblp.org}
}

@inproceedings{grain,
  author       = {Maria Drencheva and
                  Ivo Petrov and
                  Maximilian Baader and
                  Dimitar Iliev Dimitrov and
                  Martin T. Vechev},
  title        = {{GRAIN:} Exact Graph Reconstruction from Gradients},
  booktitle    = {The Thirteenth International Conference on Learning Representations,
                  {ICLR} 2025, Singapore, April 24-28, 2025},
  publisher    = {OpenReview.net},
  year         = {2025},
  url          = {https://openreview.net/forum?id=7bAjVh3CG3},
  bibsource    = {dblp computer science bibliography, https://dblp.org}
}

@inproceedings{spear,
  author       = {Dimitar I. Dimitrov and
                  Maximilian Baader and
                  Mark Niklas M{\"{u}}ller and
                  Martin T. Vechev},
  editor       = {Amir Globersons and
                  Lester Mackey and
                  Danielle Belgrave and
                  Angela Fan and
                  Ulrich Paquet and
                  Jakub M. Tomczak and
                  Cheng Zhang},
  title        = {{SPEAR:} Exact Gradient Inversion of Batches in Federated Learning},
  booktitle    = {Advances in Neural Information Processing Systems 37: Annual Conference
                  on Neural Information Processing Systems 2024, NeurIPS 2024, Vancouver,
                  BC, Canada, December 10 - 15, 2024},
  year         = {2024},
  url          = {http://papers.nips.cc/paper\_files/paper/2024/hash/c13cd7feab4beb1a27981e19e2455916-Abstract-Conference.html},
  bibsource    = {dblp computer science bibliography, https://dblp.org}
}

@article{spear++,
  author       = {Alexander Bakarsky and
                  Dimitar I. Dimitrov and
                  Maximilian Baader and
                  Martin T. Vechev},
  title        = {{SPEAR++:} Scaling Gradient Inversion via Sparsely-Used Dictionary
                  Learning},
  journal      = {CoRR},
  volume       = {abs/2510.24200},
  year         = {2025},
  url          = {https://doi.org/10.48550/arXiv.2510.24200},
  doi          = {10.48550/ARXIV.2510.24200},
  eprinttype   = {arXiv},
  eprint       = {2510.24200},
  bibsource    = {dblp computer science bibliography, https://dblp.org}
}

@inproceedings{federated,
  author       = {Brendan McMahan and
                  Eider Moore and
                  Daniel Ramage and
                  Seth Hampson and
                  Blaise Ag{\"{u}}era y Arcas},
  editor       = {Aarti Singh and
                  Xiaojin (Jerry) Zhu},
  title        = {Communication-Efficient Learning of Deep Networks from Decentralized
                  Data},
  booktitle    = {Proceedings of the 20th International Conference on Artificial Intelligence
                  and Statistics, {AISTATS} 2017, 20-22 April 2017, Fort Lauderdale,
                  FL, {USA}},
  series       = {Proceedings of Machine Learning Research},
  pages        = {1273--1282},
  publisher    = {{PMLR}},
  year         = {2017},
  url          = {http://proceedings.mlr.press/v54/mcmahan17a.html},
  bibsource    = {dblp computer science bibliography, https://dblp.org}
}

@inproceedings{april,
  author       = {Jiahao Lu and
                  Xi Sheryl Zhang and
                  Tianli Zhao and
                  Xiangyu He and
                  Jian Cheng},
  title        = {{APRIL:} Finding the Achilles' Heel on Privacy for Vision Transformers},
  booktitle    = {{IEEE/CVF} Conference on Computer Vision and Pattern Recognition,
                  {CVPR} 2022, New Orleans, LA, USA, June 18-24, 2022},
  pages        = {10041--10050},
  publisher    = {{IEEE}},
  year         = {2022},
  url          = {https://doi.org/10.1109/CVPR52688.2022.00981},
  doi          = {10.1109/CVPR52688.2022.00981},
  bibsource    = {dblp computer science bibliography, https://dblp.org}
}

@article{flowertune,
  author       = {Yan Gao and
                  Massimo Roberto Scamarcia and
                  Javier Fern{\'{a}}ndez{-}Marqu{\'{e}}s and
                  Mohammad Naseri and
                  Chong Shen Ng and
                  Dimitris Stripelis and
                  Zexi Li and
                  Tao Shen and
                  Jiamu Bai and
                  Daoyuan Chen and
                  Zikai Zhang and
                  Rui Hu and
                  Inseo Song and
                  KangYoon Lee and
                  Hong Jia and
                  Ting Dang and
                  Junyan Wang and
                  Zheyuan Liu and
                  Daniel J. Beutel and
                  Lingjuan Lyu and
                  Nicholas D. Lane},
  title        = {FlowerTune: {A} Cross-Domain Benchmark for Federated Fine-Tuning of
                  Large Language Models},
  journal      = {CoRR},
  volume       = {abs/2506.02961},
  year         = {2025},
  url          = {https://doi.org/10.48550/arXiv.2506.02961},
  doi          = {10.48550/ARXIV.2506.02961},
  eprinttype   = {arXiv},
  eprint       = {2506.02961},
  bibsource    = {dblp computer science bibliography, https://dblp.org}
}

@article{fatellm,
  author       = {Tao Fan and
                  Yan Kang and
                  Guoqiang Ma and
                  Weijing Chen and
                  Wenbin Wei and
                  Lixin Fan and
                  Qiang Yang},
  title        = {{FATE-LLM:} {A} Industrial Grade Federated Learning Framework for
                  Large Language Models},
  journal      = {CoRR},
  volume       = {abs/2310.10049},
  year         = {2023},
  url          = {https://doi.org/10.48550/arXiv.2310.10049},
  doi          = {10.48550/ARXIV.2310.10049},
  eprinttype   = {arXiv},
  eprint       = {2310.10049},
  bibsource    = {dblp computer science bibliography, https://dblp.org}
}

@inproceedings{byrddp,
  author       = {David Byrd and
                  Antigoni Polychroniadou},
  editor       = {Tucker Balch},
  title        = {Differentially private secure multi-party computation for federated
                  learning in financial applications},
  booktitle    = {{ICAIF} '20: The First {ACM} International Conference on {AI} in Finance,
                  New York, NY, USA, October 15-16, 2020},
  pages        = {16:1--16:9},
  publisher    = {{ACM}},
  year         = {2020},
  url          = {https://doi.org/10.1145/3383455.3422562},
  doi          = {10.1145/3383455.3422562},
  bibsource    = {dblp computer science bibliography, https://dblp.org}
}

@inproceedings{fedjudge,
  author       = {Linan Yue and
                  Qi Liu and
                  Yichao Du and
                  Weibo Gao and
                  Ye Liu and
                  Fangzhou Yao},
  editor       = {Makoto Onizuka and
                  Jae{-}Gil Lee and
                  Yongxin Tong and
                  Chuan Xiao and
                  Yoshiharu Ishikawa and
                  Sihem Amer{-}Yahia and
                  H. V. Jagadish and
                  Kejing Lu},
  title        = {FedJudge: Federated Legal Large Language Model},
  booktitle    = {Database Systems for Advanced Applications - 29th International Conference,
                  {DASFAA} 2024, Gifu, Japan, July 2-5, 2024, Proceedings, Part {V}},
  series       = {Lecture Notes in Computer Science},
  pages        = {268--285},
  publisher    = {Springer},
  year         = {2024},
  url          = {https://doi.org/10.1007/978-981-97-5569-1\_17},
  doi          = {10.1007/978-981-97-5569-1\_17},
  bibsource    = {dblp computer science bibliography, https://dblp.org}
}

@inproceedings{fedlegal,
  author       = {Zhuo Zhang and
                  Xiangjing Hu and
                  Jingyuan Zhang and
                  Yating Zhang and
                  Hui Wang and
                  Lizhen Qu and
                  Zenglin Xu},
  editor       = {Anna Rogers and
                  Jordan L. Boyd{-}Graber and
                  Naoaki Okazaki},
  title        = {{FEDLEGAL:} The First Real-World Federated Learning Benchmark for
                  Legal {NLP}},
  booktitle    = {Proceedings of the 61st Annual Meeting of the Association for Computational
                  Linguistics (Volume 1: Long Papers), {ACL} 2023, Toronto, Canada,
                  July 9-14, 2023},
  pages        = {3492--3507},
  publisher    = {Association for Computational Linguistics},
  year         = {2023},
  url          = {https://doi.org/10.18653/v1/2023.acl-long.193},
  doi          = {10.18653/V1/2023.ACL-LONG.193},
  bibsource    = {dblp computer science bibliography, https://dblp.org}
}

@inproceedings{manoelmedical,
  author       = {Andre Manoel and
                  Mirian del Carmen Hipolito Garcia and
                  Tal Baumel and
                  Shize Su and
                  Jialei Chen and
                  Robert Sim and
                  Dan Miller and
                  Danny Karmon and
                  Dimitrios Dimitriadis},
  editor       = {Bobak J. Mortazavi and
                  Tasmie Sarker and
                  Andrew Beam and
                  Joyce C. Ho},
  title        = {Federated Multilingual Models for Medical Transcript Analysis},
  booktitle    = {Conference on Health, Inference, and Learning, {CHIL} 2023, Broad
                  Institute of {MIT} and Harvard (Merkin Building), 415 Main Street,
                  Cambridge, MA, {USA}},
  series       = {Proceedings of Machine Learning Research},
  pages        = {147--162},
  publisher    = {{PMLR}},
  year         = {2023},
  url          = {https://proceedings.mlr.press/v209/manoel23a.html},
  bibsource    = {dblp computer science bibliography, https://dblp.org}
}

@article{zhanghealthcare,
  author       = {Lihong Zhang and
                  Yue Li},
  title        = {Federated Learning with Layer Skipping: Efficient Training of Large
                  Language Models for Healthcare {NLP}},
  journal      = {CoRR},
  volume       = {abs/2504.10536},
  year         = {2025},
  url          = {https://doi.org/10.48550/arXiv.2504.10536},
  doi          = {10.48550/ARXIV.2504.10536},
  eprinttype   = {arXiv},
  eprint       = {2504.10536},
  bibsource    = {dblp computer science bibliography, https://dblp.org}
}

@article{fedmedlora,
  author       = {Anran Li and
                  Yuanyuan Chen and
                  Wenjun Long and
                  Yu Yin and
                  Yan Hu and
                  Hyunjae Kim and
                  Weipeng Zhou and
                  Yujia Zhou and
                  Hongyi Peng and
                  Yang Ren and
                  Xuguang Ai and
                  Zhenyue Qin and
                  Ming Hu and
                  Xiaoxiao Li and
                  Han Yu and
                  Yih{-}Chung Tham and
                  Lucila Ohno{-}Machado and
                  Hua Xu and
                  Qingyu Chen},
  title        = {A Federated and Parameter-Efficient Framework for Large Language Model
                  Training in Medicine},
  journal      = {CoRR},
  volume       = {abs/2601.22124},
  year         = {2026},
  url          = {https://doi.org/10.48550/arXiv.2601.22124},
  doi          = {10.48550/ARXIV.2601.22124},
  eprinttype   = {arXiv},
  eprint       = {2601.22124},
  bibsource    = {dblp computer science bibliography, https://dblp.org}
}

@article{sadilekhealth,
  author       = {Adam Sadilek and
                  Luyang Liu and
                  Dung Nguyen and
                  Methun Kamruzzaman and
                  Stylianos Serghiou and
                  Benjamin Rader and
                  Alex Ingerman and
                  Stefan Mellem and
                  Peter Kairouz and
                  Elaine O. Nsoesie and
                  Jamie Macfarlane and
                  Anil Vullikanti and
                  Madhav V. Marathe and
                  Paul Eastham and
                  John S. Brownstein and
                  Blaise Ag{\"{u}}era y Arcas and
                  Michael D. Howell and
                  John Hernandez},
  title        = {Privacy-first health research with federated learning},
  journal      = {npj Digit. Medicine},
  volume       = {4},
  year         = {2021},
  url          = {https://doi.org/10.1038/s41746-021-00489-2},
  doi          = {10.1038/S41746-021-00489-2},
  bibsource    = {dblp computer science bibliography, https://dblp.org}
}

@article{chenfederated,
  author       = {Chaochao Chen and
                  Xiaohua Feng and
                  Yuyuan Li and
                  Lingjuan Lyu and
                  Jun Zhou and
                  Xiaolin Zheng and
                  Jianwei Yin},
  title        = {Integration of large language models and federated learning},
  journal      = {Patterns},
  volume       = {5},
  number       = {12},
  pages        = {101098},
  year         = {2024},
  url          = {https://doi.org/10.1016/j.patter.2024.101098},
  doi          = {10.1016/J.PATTER.2024.101098},
  bibsource    = {dblp computer science bibliography, https://dblp.org}
}

@article{yaofederatedllm,
  author       = {Yuhang Yao and
                  Jianyi Zhang and
                  Junda Wu and
                  Chengkai Huang and
                  Yu Xia and
                  Tong Yu and
                  Ruiyi Zhang and
                  Sungchul Kim and
                  Ryan A. Rossi and
                  Ang Li and
                  Lina Yao and
                  Julian J. McAuley and
                  Yiran Chen and
                  Carlee Joe{-}Wong},
  title        = {Federated Large Language Models: Current Progress and Future Directions},
  journal      = {CoRR},
  volume       = {abs/2409.15723},
  year         = {2024},
  url          = {https://doi.org/10.48550/arXiv.2409.15723},
  doi          = {10.48550/ARXIV.2409.15723},
  eprinttype   = {arXiv},
  eprint       = {2409.15723},
  bibsource    = {dblp computer science bibliography, https://dblp.org}
}

@article{jiangfederatedllm,
  author       = {Wenhao Jiang and
                  Yuchuan Luo and
                  Guilin Deng and
                  Silong Chen and
                  Xu Yang and
                  Shihong Wu and
                  Xinwen Gao and
                  Lin Liu and
                  Shaojing Fu},
  title        = {Federated Large Language Models: Feasibility, Robustness, Security
                  and Future Directions},
  journal      = {CoRR},
  volume       = {abs/2505.08830},
  year         = {2025},
  url          = {https://doi.org/10.48550/arXiv.2505.08830},
  doi          = {10.48550/ARXIV.2505.08830},
  eprinttype   = {arXiv},
  eprint       = {2505.08830},
  bibsource    = {dblp computer science bibliography, https://dblp.org}
}

@inproceedings{film,
  author       = {Samyak Gupta and
                  Yangsibo Huang and
                  Zexuan Zhong and
                  Tianyu Gao and
                  Kai Li and
                  Danqi Chen},
  editor       = {Sanmi Koyejo and
                  S. Mohamed and
                  A. Agarwal and
                  Danielle Belgrave and
                  K. Cho and
                  A. Oh},
  title        = {Recovering Private Text in Federated Learning of Language Models},
  booktitle    = {Advances in Neural Information Processing Systems 35: Annual Conference
                  on Neural Information Processing Systems 2022, NeurIPS 2022, New Orleans,
                  LA, USA, November 28 - December 9, 2022},
  year         = {2022},
  url          = {http://papers.nips.cc/paper\_files/paper/2022/hash/35b5c175e139bff5f22a5361270fce87-Abstract-Conference.html},
  bibsource    = {dblp computer science bibliography, https://dblp.org}
}

@article{poolerattack,
  author       = {Jianwei Li and
                  Sheng Liu and
                  Qi Lei},
  title        = {Beyond Gradient and Priors in Privacy Attacks: Leveraging Pooler Layer
                  Inputs of Language Models in Federated Learning},
  journal      = {CoRR},
  volume       = {abs/2312.05720},
  year         = {2023},
  url          = {https://doi.org/10.48550/arXiv.2312.05720},
  doi          = {10.48550/ARXIV.2312.05720},
  eprinttype   = {arXiv},
  eprint       = {2312.05720},
  bibsource    = {dblp computer science bibliography, https://dblp.org}
}

@inproceedings{robbingthefed,
  author       = {Liam H. Fowl and
                  Jonas Geiping and
                  Wojciech Czaja and
                  Micah Goldblum and
                  Tom Goldstein},
  title        = {Robbing the Fed: Directly Obtaining Private Data in Federated Learning
                  with Modified Models},
  booktitle    = {The Tenth International Conference on Learning Representations, {ICLR}
                  2022, Virtual Event, April 25-29, 2022},
  publisher    = {OpenReview.net},
  year         = {2022},
  url          = {https://openreview.net/forum?id=fwzUgo0FM9v},
  bibsource    = {dblp computer science bibliography, https://dblp.org}
}

@inproceedings{fishing,
  author       = {Yuxin Wen and
                  Jonas Geiping and
                  Liam Fowl and
                  Micah Goldblum and
                  Tom Goldstein},
  editor       = {Kamalika Chaudhuri and
                  Stefanie Jegelka and
                  Le Song and
                  Csaba Szepesv{\'{a}}ri and
                  Gang Niu and
                  Sivan Sabato},
  title        = {Fishing for User Data in Large-Batch Federated Learning via Gradient
                  Magnification},
  booktitle    = {International Conference on Machine Learning, {ICML} 2022, 17-23 July
                  2022, Baltimore, Maryland, {USA}},
  series       = {Proceedings of Machine Learning Research},
  pages        = {23668--23684},
  publisher    = {{PMLR}},
  year         = {2022},
  url          = {https://proceedings.mlr.press/v162/wen22a.html},
  bibsource    = {dblp computer science bibliography, https://dblp.org}
}

@inproceedings{decepticons,
  author       = {Liam H. Fowl and
                  Jonas Geiping and
                  Steven Reich and
                  Yuxin Wen and
                  Wojciech Czaja and
                  Micah Goldblum and
                  Tom Goldstein},
  title        = {Decepticons: Corrupted Transformers Breach Privacy in Federated Learning
                  for Language Models},
  booktitle    = {The Eleventh International Conference on Learning Representations,
                  {ICLR} 2023, Kigali, Rwanda, May 1-5, 2023},
  publisher    = {OpenReview.net},
  year         = {2023},
  url          = {https://openreview.net/forum?id=r0BrY4BiEXO},
  bibsource    = {dblp computer science bibliography, https://dblp.org}
}

@inproceedings{panningforgold,
  author       = {Hong{-}Min Chu and
                  Jonas Geiping and
                  Liam H. Fowl and
                  Micah Goldblum and
                  Tom Goldstein},
  title        = {Panning for Gold in Federated Learning: Targeted Text Extraction under
                  Arbitrarily Large-Scale Aggregation},
  booktitle    = {The Eleventh International Conference on Learning Representations,
                  {ICLR} 2023, Kigali, Rwanda, May 1-5, 2023},
  publisher    = {OpenReview.net},
  year         = {2023},
  url          = {https://openreview.net/forum?id=A9WQaxYsfx},
  bibsource    = {dblp computer science bibliography, https://dblp.org}
}

@inproceedings{
minegrad,
title={MineGrad: Gradient Inversion Attacks on Lo{RA} Fine-Tuning},
author={Hasin Us Sami and Swapneel Sen and Basak Guler},
booktitle={The 29th International Conference on Artificial Intelligence and Statistics},
year={2026},
url={https://openreview.net/forum?id=dD9XOZUpNc}
}

@article{idlg,
  author       = {Bo Zhao and
                  Konda Reddy Mopuri and
                  Hakan Bilen},
  title        = {iDLG: Improved Deep Leakage from Gradients},
  journal      = {CoRR},
  volume       = {abs/2001.02610},
  year         = {2020},
  url          = {http://arxiv.org/abs/2001.02610},
  eprinttype   = {arXiv},
  eprint       = {2001.02610},
  bibsource    = {dblp computer science bibliography, https://dblp.org}
}

@article{towardsgeneral,
  author       = {Jiahui Geng and
                  Yongli Mou and
                  Feifei Li and
                  Qing Li and
                  Oya Beyan and
                  Stefan Decker and
                  Chunming Rong},
  title        = {Towards General Deep Leakage in Federated Learning},
  journal      = {CoRR},
  volume       = {abs/2110.09074},
  year         = {2021},
  url          = {https://arxiv.org/abs/2110.09074},
  eprinttype   = {arXiv},
  eprint       = {2110.09074},
  bibsource    = {dblp computer science bibliography, https://dblp.org}
}

@inproceedings{rgap,
  author       = {Junyi Zhu and
                  Matthew B. Blaschko},
  title        = {{R-GAP:} Recursive Gradient Attack on Privacy},
  booktitle    = {9th International Conference on Learning Representations, {ICLR} 2021,
                  Virtual Event, Austria, May 3-7, 2021},
  publisher    = {OpenReview.net},
  year         = {2021},
  url          = {https://openreview.net/forum?id=RSU17UoKfJF},
  bibsource    = {dblp computer science bibliography, https://dblp.org}
}

@inproceedings{dpsgd,
  author       = {Mart{\'{\i}}n Abadi and
                  Andy Chu and
                  Ian J. Goodfellow and
                  H. Brendan McMahan and
                  Ilya Mironov and
                  Kunal Talwar and
                  Li Zhang},
  editor       = {Edgar R. Weippl and
                  Stefan Katzenbeisser and
                  Christopher Kruegel and
                  Andrew C. Myers and
                  Shai Halevi},
  title        = {Deep Learning with Differential Privacy},
  booktitle    = {Proceedings of the 2016 {ACM} {SIGSAC} Conference on Computer and
                  Communications Security, Vienna, Austria, October 24-28, 2016},
  pages        = {308--318},
  publisher    = {{ACM}},
  year         = {2016},
  url          = {https://doi.org/10.1145/2976749.2978318},
  doi          = {10.1145/2976749.2978318},
  bibsource    = {dblp computer science bibliography, https://dblp.org}
}

@article{fictionalqa,
  author       = {John Kirchenbauer and
                  Janny Mongkolsupawan and
                  Yuxin Wen and
                  Tom Goldstein and
                  Daphne Ippolito},
  title        = {A Fictional Q{\&}A Dataset for Studying Memorization and Knowledge
                  Acquisition},
  journal      = {CoRR},
  volume       = {abs/2506.05639},
  year         = {2025},
  url          = {https://doi.org/10.48550/arXiv.2506.05639},
  doi          = {10.48550/ARXIV.2506.05639},
  eprinttype   = {arXiv},
  eprint       = {2506.05639},
  bibsource    = {dblp computer science bibliography, https://dblp.org}
}

@article{clipping,
  author       = {Robin C. Geyer and
                  Tassilo Klein and
                  Moin Nabi},
  title        = {Differentially Private Federated Learning: {A} Client Level Perspective},
  journal      = {CoRR},
  volume       = {abs/1712.07557},
  year         = {2017},
  url          = {http://arxiv.org/abs/1712.07557},
  eprinttype   = {arXiv},
  eprint       = {1712.07557},
  bibsource    = {dblp computer science bibliography, https://dblp.org}
}

@inproceedings{adaptiveclipping,
  author       = {Galen Andrew and
                  Om Thakkar and
                  Brendan McMahan and
                  Swaroop Ramaswamy},
  editor       = {Marc'Aurelio Ranzato and
                  Alina Beygelzimer and
                  Yann N. Dauphin and
                  Percy Liang and
                  Jennifer Wortman Vaughan},
  title        = {Differentially Private Learning with Adaptive Clipping},
  booktitle    = {Advances in Neural Information Processing Systems 34: Annual Conference
                  on Neural Information Processing Systems 2021, NeurIPS 2021, December
                  6-14, 2021, virtual},
  pages        = {17455--17466},
  year         = {2021},
  url          = {https://proceedings.neurips.cc/paper/2021/hash/91cff01af640a24e7f9f7a5ab407889f-Abstract.html},
  bibsource    = {dblp computer science bibliography, https://dblp.org}
}

@inproceedings{sampleadaptiveclipping,
  author       = {Tianyu Xia and
                  Shuheng Shen and
                  Su Yao and
                  Xinyi Fu and
                  Ke Xu and
                  Xiaolong Xu and
                  Xing Fu},
  editor       = {Brian Williams and
                  Yiling Chen and
                  Jennifer Neville},
  title        = {Differentially Private Learning with Per-Sample Adaptive Clipping},
  booktitle    = {Thirty-Seventh {AAAI} Conference on Artificial Intelligence, {AAAI}
                  2023, Thirty-Fifth Conference on Innovative Applications of Artificial
                  Intelligence, {IAAI} 2023, Thirteenth Symposium on Educational Advances
                  in Artificial Intelligence, {EAAI} 2023, Washington, DC, USA, February
                  7-14, 2023},
  pages        = {10444--10452},
  publisher    = {{AAAI} Press},
  year         = {2023},
  url          = {https://doi.org/10.1609/aaai.v37i9.26242},
  doi          = {10.1609/AAAI.V37I9.26242},
  bibsource    = {dblp computer science bibliography, https://dblp.org}
}

@inproceedings{secureaggregation,
  author       = {Kallista A. Bonawitz and
                  Vladimir Ivanov and
                  Ben Kreuter and
                  Antonio Marcedone and
                  H. Brendan McMahan and
                  Sarvar Patel and
                  Daniel Ramage and
                  Aaron Segal and
                  Karn Seth},
  editor       = {Bhavani Thuraisingham and
                  David Evans and
                  Tal Malkin and
                  Dongyan Xu},
  title        = {Practical Secure Aggregation for Privacy-Preserving Machine Learning},
  booktitle    = {Proceedings of the 2017 {ACM} {SIGSAC} Conference on Computer and
                  Communications Security, {CCS} 2017, Dallas, TX, USA, October 30 -
                  November 03, 2017},
  pages        = {1175--1191},
  publisher    = {{ACM}},
  year         = {2017},
  url          = {https://doi.org/10.1145/3133956.3133982},
  doi          = {10.1145/3133956.3133982},
  bibsource    = {dblp computer science bibliography, https://dblp.org}
}

@inproceedings{compression,
  author       = {Yujun Lin and
                  Song Han and
                  Huizi Mao and
                  Yu Wang and
                  Bill Dally},
  title        = {Deep Gradient Compression: Reducing the Communication Bandwidth for
                  Distributed Training},
  booktitle    = {6th International Conference on Learning Representations, {ICLR} 2018,
                  Vancouver, BC, Canada, April 30 - May 3, 2018, Conference Track Proceedings},
  publisher    = {OpenReview.net},
  year         = {2018},
  url          = {https://openreview.net/forum?id=SkhQHMW0W},
  bibsource    = {dblp computer science bibliography, https://dblp.org}
}

@article{communicationstrategies,
  author       = {Jakub Kone{\v{c}}n{\'y} and
                  H. Brendan McMahan and
                  Felix X. Yu and
                  Peter Richt{\'{a}}rik and
                  Ananda Theertha Suresh and
                  Dave Bacon},
  title        = {Federated Learning: Strategies for Improving Communication Efficiency},
  journal      = {CoRR},
  volume       = {abs/1610.05492},
  year         = {2016},
  url          = {http://arxiv.org/abs/1610.05492},
  eprinttype   = {arXiv},
  eprint       = {1610.05492},
  bibsource    = {dblp computer science bibliography, https://dblp.org}
}

@inproceedings{rouge,
    title = "{ROUGE}: A Package for Automatic Evaluation of Summaries",
    author = "Lin, Chin-Yew",
    booktitle = "Text Summarization Branches Out",
    month = jul,
    year = "2004",
    address = "Barcelona, Spain",
    publisher = "Association for Computational Linguistics",
    url = "https://aclanthology.org/W04-1013/",
    pages = "74--81"
}

@article{embeddinggemma,
    title={EmbeddingGemma: Powerful and Lightweight Text Representations},
    publisher={Google DeepMind},
    author={Schechter Vera, Henrique and Dua, Sahil and {EmbeddingGemma Team}},
    year={2025},
    url={https://arxiv.org/abs/2509.20354}
}

@article{gemma3,
    title={Gemma 3},
    url={https://arxiv.org/abs/2503.19786},
    publisher={Google DeepMind},
    author={{Gemma Team}},
    year={2025}
}

@article{medgemma,
  author       = {{Google Research} and
                  {Google DeepMind}},
  title        = {MedGemma Technical Report},
  journal      = {CoRR},
  volume       = {abs/2507.05201},
  year         = {2025},
  url          = {https://doi.org/10.48550/arXiv.2507.05201},
  doi          = {10.48550/ARXIV.2507.05201},
  eprinttype   = {arXiv},
  eprint       = {2507.05201},
  bibsource    = {dblp computer science bibliography, https://dblp.org}
}

@article{medgemma15,
  author       = {Andrew Sellergren and
                  Chufan Gao and
                  Fereshteh Mahvar and
                  Timo Kohlberger and
                  Fayaz Jamil and
                  Madeleine Traverse and
                  Alberto Tono and
                  Bashir Sadjad and
                  Lin Yang and
                  Charles Lau and
                  Liron Yatziv and
                  Tiffany L. Chen and
                  Bram Sterling and
                  Kenneth Philbrick and
                  Richa Tiwari and
                  Yun Liu and
                  Madhuram Jajoo and
                  Chandrashekar Sankarapu and
                  Swapnil Vispute and
                  Harshad Purandare and
                  Abhishek Bijay Mishra and
                  Samuel Schmidgall and
                  Tao Tu and
                  Anil Palepu and
                  Chunjong Park and
                  Tim Strother and
                  Rahul Thapa and
                  Yong Cheng and
                  Preeti Singh and
                  Kat Black and
                  Yossi Matias and
                  Katherine Chou and
                  Avinatan Hassidim and
                  Kavi Goel and
                  Joelle K. Barral and
                  Tris Warkentin and
                  Shravya Shetty and
                  Dale R. Webster and
                  Sunny Virmani and
                  David F. Steiner and
                  Can Kirmizibayrak and
                  Daniel Golden},
  title        = {MedGemma 1.5 Technical Report},
  journal      = {CoRR},
  volume       = {abs/2604.05081},
  year         = {2026},
  url          = {https://doi.org/10.48550/arXiv.2604.05081},
  doi          = {10.48550/ARXIV.2604.05081},
  eprinttype   = {arXiv},
  eprint       = {2604.05081},
  bibsource    = {dblp computer science bibliography, https://dblp.org}
}

@misc{embeddinggemmaft,
  title  = {An Extended Annotation Scheme for Personal-Fact Classification in Dialogue},
  author = {Zaitsev, Konstantin},
  year   = {2026},
  note   = {Model: https://huggingface.co/adugeen/personal-facts-classifier-embeddinggemma-300m;
            Dataset: https://huggingface.co/datasets/adugeen/personal-facts-msc}
}

@misc{financialgemma,
      title={How Small Can You Go? LoRA Fine-Tuning 270M-8B Models for Merchant Information Extraction in Financial Transactions}, 
      author={Donghao Huang and Tomas Drietomsky and Benjamin Barrett and Zhaoxia Wang},
      year={2026},
      eprint={2606.08051},
      archivePrefix={arXiv},
      primaryClass={cs.AI},
      url={https://arxiv.org/abs/2606.08051}, 
}

@inproceedings{gradinversion,
  author       = {Hongxu Yin and
                  Arun Mallya and
                  Arash Vahdat and
                  Jos{\'{e}} M. {\'{A}}lvarez and
                  Jan Kautz and
                  Pavlo Molchanov},
  title        = {See Through Gradients: Image Batch Recovery via GradInversion},
  booktitle    = {{IEEE} Conference on Computer Vision and Pattern Recognition, {CVPR}
                  2021, virtual, June 19-25, 2021},
  pages        = {16337--16346},
  publisher    = {Computer Vision Foundation / {IEEE}},
  year         = {2021},
  url          = {https://openaccess.thecvf.com/content/CVPR2021/html/Yin\_See\_Through\_Gradients\_Image\_Batch\_Recovery\_via\_GradInversion\_CVPR\_2021\_paper.html},
  doi          = {10.1109/CVPR46437.2021.01607},
  bibsource    = {dblp computer science bibliography, https://dblp.org}
}

@inproceedings{generativegradinversion,
  author       = {Jinwoo Jeon and
                  Jaechang Kim and
                  Kangwook Lee and
                  Sewoong Oh and
                  Jungseul Ok},
  editor       = {Marc'Aurelio Ranzato and
                  Alina Beygelzimer and
                  Yann N. Dauphin and
                  Percy Liang and
                  Jennifer Wortman Vaughan},
  title        = {Gradient Inversion with Generative Image Prior},
  booktitle    = {Advances in Neural Information Processing Systems 34: Annual Conference
                  on Neural Information Processing Systems 2021, NeurIPS 2021, December
                  6-14, 2021, virtual},
  pages        = {29898--29908},
  year         = {2021},
  url          = {https://proceedings.neurips.cc/paper/2021/hash/fa84632d742f2729dc32ce8cb5d49733-Abstract.html},
  bibsource    = {dblp computer science bibliography, https://dblp.org}
}

@inproceedings{gradvit,
  author       = {Ali Hatamizadeh and
                  Hongxu Yin and
                  Holger Roth and
                  Wenqi Li and
                  Jan Kautz and
                  Daguang Xu and
                  Pavlo Molchanov},
  title        = {GradViT: Gradient Inversion of Vision Transformers},
  booktitle    = {{IEEE/CVF} Conference on Computer Vision and Pattern Recognition,
                  {CVPR} 2022, New Orleans, LA, USA, June 18-24, 2022},
  pages        = {10011--10020},
  publisher    = {{IEEE}},
  year         = {2022},
  url          = {https://doi.org/10.1109/CVPR52688.2022.00978},
  doi          = {10.1109/CVPR52688.2022.00978},
  bibsource    = {dblp computer science bibliography, https://dblp.org}
}

@inproceedings{cocktailparty,
  author       = {Sanjay Kariyappa and
                  Chuan Guo and
                  Kiwan Maeng and
                  Wenjie Xiong and
                  G. Edward Suh and
                  Moinuddin K. Qureshi and
                  Hsien{-}Hsin S. Lee},
  editor       = {Andreas Krause and
                  Emma Brunskill and
                  Kyunghyun Cho and
                  Barbara Engelhardt and
                  Sivan Sabato and
                  Jonathan Scarlett},
  title        = {Cocktail Party Attack: Breaking Aggregation-Based Privacy in Federated
                  Learning Using Independent Component Analysis},
  booktitle    = {International Conference on Machine Learning, {ICML} 2023, 23-29 July
                  2023, Honolulu, Hawaii, {USA}},
  series       = {Proceedings of Machine Learning Research},
  pages        = {15884--15899},
  publisher    = {{PMLR}},
  year         = {2023},
  url          = {https://proceedings.mlr.press/v202/kariyappa23a.html},
  bibsource    = {dblp computer science bibliography, https://dblp.org}
}

@misc{wikitext,
      title={Pointer Sentinel Mixture Models},
      author={Stephen Merity and Caiming Xiong and James Bradbury and Richard Socher},
      year={2016},
      eprint={1609.07843},
      archivePrefix={arXiv},
      primaryClass={cs.CL}
}

@InProceedings{imdb,
  author    = {Maas, Andrew L.  and  Daly, Raymond E.  and  Pham, Peter T.  and  Huang, Dan  and  Ng, Andrew Y.  and  Potts, Christopher},
  title     = {Learning Word Vectors for Sentiment Analysis},
  booktitle = {Proceedings of the 49th Annual Meeting of the Association for Computational Linguistics: Human Language Technologies},
  month     = {June},
  year      = {2011},
  address   = {Portland, Oregon, USA},
  publisher = {Association for Computational Linguistics},
  pages     = {142--150},
  url       = {http://www.aclweb.org/anthology/P11-1015}
}
